\documentclass[apj]{emulateapj}

\usepackage{graphicx}
\usepackage[space]{grffile}
\usepackage{latexsym}
\usepackage{amsfonts,amsmath,amssymb}
\usepackage{url}
\usepackage[utf8]{inputenc}
\usepackage{fancyref}
\usepackage{hyperref}
\hypersetup{colorlinks=false,pdfborder={0 0 0},}
\usepackage{epstopdf}
\usepackage{natbib}

\slugcomment{To appear in The Astrophysical Journal (ApJ)}

\shorttitle{Astrochemical Correlations in Molecular Clouds}
\shortauthors{Gaches et al.}

\begin{document}

\title{Astrochemical Correlations in Molecular Clouds}

\author{Brandt A.L. Gaches}
\affil{Department of Astronomy, University of Massachusetts - Amherst}
\email{bgaches@astro.umass.edu}

\author{Stella S.R. Offner\altaffilmark{1,2}}
\affil{Department of Astronomy, University of Massachusetts - Amherst}
\email{soffner@astro.umass.edu}

\author{Erik W. Rosolowsky}
\affil{Department of Physics, University of Alberta}
\email{erosolow@ualberta.ca}

\author{Thomas G. Bisbas}
\affil{Department of Physics and Astronomy, University College London}
\email{tb@star.ucl.ac.uk}

\altaffiltext{1}{Hubble Fellow}
\altaffiltext{2}{Department of Astronomy, Yale University}

\begin{abstract}
We investigate the spectral correlations between different species
used to observe molecular clouds. We use hydrodynamic
simulations and a full chemical network to study the abundances of over 150 species in typical Milky Way molecular clouds. We perform synthetic observations in order to produce emission maps of a subset of these tracers. We study the effects of different lines of sight and spatial resolution on the emission distribution and perform a robust quantitative comparison of the species to each other. We use the Spectral Correlation Function (SCF), which quantifies the root mean squared difference between spectra separated by some length scale, to characterize the structure of the simulated cloud in position-position-velocity (PPV) space. We predict the observed SCF for a broad range of observational tracers, and thus, identify homologous species. In particular, we show that the pairs C and CO, C$^{+}$ and CN, NH$_3$ and H$_2$CS have very similar SCFs. We measure the SCF slope variation as a function of beam size for all species and demonstrate that the beam size has a distinct effect on different species emission. However, for beams of up to 10'', placing the cloud at 1 kpc, the change is not large enough to move the SCF slopes into different regions of parameter space. The results from this study provide observational guidance for choosing the best tracer to probe various cloud length scales.

\end{abstract}

\keywords{astrochemistry, ISM:clouds, ISM:molecules, ISM:structure, methods:numerical, turbulence}

\section{Introduction}

In molecular clouds, the largest component of the mass is in the form of molecular hydrogen. However, the lowest transition of H$_2$ is excited at temperatures greater than 500 K, an order of magnitude hotter than typical molecular clouds (10-100 K). Therefore, observational studies of molecular cloud properties and dynamics must use low-abundance tracer species, the most important being CO \citep[see][]{2006MNRAS.371.1865B, 2011MNRAS.412..337G,2011MNRAS.412.1686S}. Despite its utility, CO is an imperfect gas tracer in a variety of regimes. It is photo-dissociated at cloud boundaries and near stellar sources. It also becomes optically thick quickly in high density regions. Furthermore, CO suffers greatly from depletion onto dust grains at gas densities greater than $\sim$ 10$^4$ cm$^{-3}$, further reducing its ability to probe high density environments \citep[][]{2002A&A...389L...6B, 2007ARA&A..45..339B, 2014MNRAS.438L..56H}. In contrast, high-density tracers such as N$_2$H$^+$ provide a better means to study the dense gas where CO depletion becomes severe.

Consequently, in order to construct a complete picture of molecular clouds, it is necessary to piece together information from a variety of different tracers and line transitions, each of which is sensitive to different densities, temperatures, and size scales \citep[e.g.][]{1998ApJ...504..223G}. For diffuse gas, species such as OH$^{+}$ \citep{2014ApJ...781L...8P}  and C$^{+}$ are used. Shock dominated regions are traced by molecules such as SiO \citep{2014arXiv1402.5066O}. Observers also use a wide array of species, including HCN, NH$_3$, H$_2$CO, H$_2$CS, CS, HCO$^{+}$ and N$_2$H$^{+}$ to study dense environments \citep{2008ApJS..175..509R, 2011MNRAS.415.1977R, 2012ApJ...756...60S, 2013arXiv1312.1905H, 2014ApJ...780...85V}. However, astrochemistry is highly nonlinear and is sensitive to the gas density, temperature and radiation field. Detailed modelling is necessary to understand the distribution of species and how they are correlated with each other.

Observationally, new instruments make studying many different chemical species much more feasible. In particlar, the wide spectral bandwidth of Atacama Large Millimeter Array (ALMA) allows many different molecular species to be be mapped simultaneously. For example, the ALMA Band 3 contains CN (N=1=0, J=$\frac{1}{2}$-$\frac{1}{2}$), $^{\rm 12}$CO (J=1-0), $^{\rm 13}$CO (J=1-0), HCN (J=1-0), HCO$^+$ (J=1-0), HNC (J=1-0), and N$_2$H$^+$ (J=1-0). These species alone span environments ranging from diffuse, large scale structure ($^{\rm 12}$CO) to dense cores (N$_2$H$^+$). Altogether, these lines provide a rich and detailed view of star forming gas.

However, interpreting this data is not always straight forward. Astrochemistry is a highly nonlinear function of the underlying gas density and temperature. Numerical modelling of the underlying structure and astrochemistry is essential to provide an interpretive framework. There have been numerous recent advancements in astrochemistry codes \citep[see][]{2007A&A...467..187R}. Reducing the dimensionality can be helpful for modelling disk or outflow cavities \citep[e.g.][]{2009ApJ...700..872B, 2009A&A...501..383W}, where the underlying density structure is symmetric is some manner. However, observations show that molecular clouds contain complex density and velocity structure. To account for this, some studies have adopted analytic prescriptions for clumpiness \citep{1997A&A...323..953S, 2008A&A...488..623C}. Such approaches have been able to model the correlation between $^{13}$CO and C, CO emission (the ``X-factor") and reproduce observed line profiles \citep[e.g.][]{2008A&A...477..547K, 2013ARA&A..51..207B}, however, they underestimate the influence of cloud morphology in the abundances of species such as C \citep{2004MNRAS.351..147P,2014arXiv1403.3530G, 2014MNRAS.tmpL..37O}. A fully three dimensional method for modelling photo-dissociation regions (PDRs) is necessary to represent the complex geometries of molecular clouds. However, a three dimensional time dependent astrochemistry simulation coupled to the hydrodynamics with an extended network is still computationally infeasible.

Many astrochemical studies have focused only on the formation of H$_2$ and CO. These species represent most of the molecular cloud gas and can be modelled with a simple chemical network \citep[see][]{2010ApJ...716.1191W}. This can reduce the problem of solving thousands of coupled equations to merely dozens. However, as more elements are added to the network, the number of species increases drastically, sometimes with significant effects \citep[][]{1999RvMP...71..173H}. For instance, the inclusion of Sulfur chemistry leads to the addition of CS, which in turn impacts the abundance of CO. Since CO and CS have similar formation rates in diffuse environments from their atomic constituents, CS will reduce some of the atomic Carbon available for the formation of CO. Likewise, CH$_3$ is a reactant with both H$_2$CS and H$_2$CO. Without the inclusion of Sulfur chemistry, the abundance of H$_2$CO cannot be properly modelled. When studying many species simultaneously, reducing the number of reactions while maintaining accuracy is nearly {\it impossible}, requiring a very large network to properly account for the gas phase chemistry alone \citep[][]{2007A&A...467..187R}. Including grain surface chemistry, which requires also modelling the dust distribution and gas-dust surface reactions, further complicates astrochemical studies.

The goal of this paper is to study the distribution of a large variety of species and understand the correlations between them. We model a Milky Way-like molecular cloud using the magnetohydrodynamic code, {\sc orion}. We then post processes these results with the 3D astrochemistry code, {\sc 3d-pdr} \citep{2012MNRAS.427.2100B}, using an extended chemical network to obtain the abundances of over 200 different species. For a subset of this network, we calculate synthetic emission maps of their lowest transition with the radiative transfer code {\sc radmc-3d} We compare the spectral structure of all the species to determine the similarity of different species. This is the first study to quantitatively compare the underlying structure of many species. In Section 2, we discuss the models and the methodology we use to generate the emission maps, as well as the way we quantify their spectral structure. In Section 3 we investigate how the following factors affect the spectral structure: viewing angle, species and spatial resolution. Finally, in Section 4 we discuss how the results are relevant to observational studies and general implications.

\section{Methods}
\subsection{Hydrodynamic Simulation}\label{sec:hydsim}

We use a turbulent hydrodynamic simulation to model a typical Milky Way cloud. The simulation was performed using the Adaptive Mesh Refinement (AMR) code {\sc orion} \citep{truelove98,klein99}. It was previously discussed in \citet{2013ApJ...770...49O} and \citet{2014MNRAS.tmpL..37O} as simulation {\it Rm6}, so we only briefly summarize its properties here. 

The simulation represents a piece of a typical local molecular cloud. The domain is 2 pc on a side and contains 600 $M_{\sun}$. The gas temperature is 10 K and the 1D gas velocity dispersion is 0.72 km s$^{-1}$ such that the cloud satisfies the linewidth-size relation (e.g., \citealt{2007ARA&A..45..565M}).  The gas turbulence is initialized by adding random velocity perturbations with wave numbers $k=1..2$ for two box crossing times without self-gravity. 

Once self-gravity is turned on collapse proceeds. The base grid is $256^3$ but four additional AMR levels are inserted to ensure that the gas obeys the Jeans criterion with a Jeans number of 0.125 \citep{truelove97}. On the highest level ($\Delta x_{\rm min}$=0.001 pc), sink particles (``stars") are added when the gas exceeds the maximum density \citep{krumholz04}. The output we analyze in this study is at 1 global free fall time at which point $\sim18\%$ of the gas resides in stars.

\subsection{Astrochemistry}\label{sec:astrochem}

The chemical abundances were computed using {\sc 3d-pdr},  a three-dimensional photodissociation code coupled with a full chemical network \citep{2012MNRAS.427.2100B}. \citet{2014MNRAS.tmpL..37O} presented an analysis of the line emission based upon the abundances of molecular hydrogen,  atomic carbon, and carbon monoxide (run Rm6\_1.0\_12\_1f\_a). In this study, we perform an additional analysis of the most common astrochemical species in the chemical network. Here, we describe the astrochemistry calculation procedure and refer the reader to \citet{2012MNRAS.427.2100B} and \citet{2014MNRAS.tmpL..37O} for additional details.

We irradiated the simulated cloud by an isotropic 1 Draine FUV field, where ``1 Draine" is the standard interstellar radiation field \citep{draine78}.  {\sc 3d-pdr} uses the hydrodynamic densities and assumed external field to compute the temperature and abundance distribution for points in the cloud with densities $200 \leq n \leq 10^5$ cm$^{-3}$. Below $n = 200$ cm$^{-3}$ we consider the gas as ionized, { using limiting conditions on abundances and gas temperature}, whereas above $n = 10^5$ cm$^{-3}$ we assume it is fully molecular. We do not calculate the chemistry in those two regimes. {\sc 3d-pdr} computes the radiation field using a resolution of 12 {\sc healpix} rays \citep{gorski05}, emanating from each grid point.  The input grid is the density field of the hydrodynamic calculation resampled to a resolution of $256^3/12$, i.e. a new grid comprised of every 12th data point. (See \citet{2013ApJ...770...49O} for a discussion of spatial resolution convergence.)

{\sc 3d-pdr} employs the UMIST2012 chemical database \citep{mcelroy13}, which includes 215 species and approximately 3000 reactions. The calculation includes the formation of H$_2$ on dust grains following \citet{cazaux02, cazaux04}, photodissociation of H$_2$ and CO and self-shielding.  The initial elemental abundances are [He] = $1.0 \times 10^{-1}$, [C] $= 1.41 \times 10^{-4}$, [O] $= 3.16 \times 10^{-4}$, [Mg] $= 5.1 \times10^{-6}$, [S] $= 1.4 \times 10^{-6}$, and [Fe] $= 3.6 \times 10^{-7}$ \citep[][]{2009ARA&A..47..481A}, which are similar to values estimated for local molecular clouds. We adopt a cosmic ionization rate of $5\times10^{-17}$ s$^{-1}$, which is the average value found in the Milky Way. {\sc 3d-pdr} solves the chemical network in equilibrium, with the final time parameter representing the time allowed for the chemistry to come to equilibrium.We analyze the calculation after advancing the chemistry to equilibrium at 10 Myr.

The {\sc 3d-pdr}  calculation does not consider shock chemistry or dust-grain chemistry, which includes the freeze-out of species such as CO onto dust grains, surface reactions (with the exception of H$_2$ formation), or release of grain mantle species into the gas phase by means of evaporation, photodesorption or desorption \citep[e.g.][]{viti04}.  While these processes may impact abundances under certain conditions, we expect them to have a minimal impact on our results. For example,  turbulent intermittency in the form of strong, thin shocks within the diffuse gas may enhance the emission and abundance of tracers such as CH$^+$,  H$_2$ and CO \citep{falgarone05,falgarone09}. However, excitation by FUV photons likely dominates the populations of the lowest energy states, which are what we study here. Dust grain freeze-out primarily affects gas with densities $\geq 10^4$ cm$^{-3}$; here, only 1\% of the volume of our model cloud has $n_{\rm H_2} \geq$ 10$^4$ cm$^{-3}$ (see further discussion of the impact of molecular freeze-out in the Appendix). Other grain processes, such as photodesorption,  may impact the species abundances over a larger range of densities. Namely, \citep{guzman13} find that photodesorption is needed to reproduce the measured enhanced abundance of H$_2$CO in the Horsehead PDR region. In contrast, they find that grain-chemistry is not required to model the H$_2$CO abundance within a UV-shielded dense core, which is instead well-fit by a pure gas-phase model. This suggests that the higher UV radiated outer regions of our cloud may require consideration of additional processes, at least with respect to H$_2$CO. However, we note that grain chemistry involves a high-degree of uncertainty and is sensitive to the local UV field, grain properties, and cosmic-ray flux, which make it difficult to apply conclusions from case-studies in the literature to our particular conditions. We quantitatively examine the impact of potential effects of dust-grain chemistry and shock chemistry in the Appendix.

In this study, we essentially assume a ``one-way" coupling between the hydrodynamics and chemistry. Performing the chemistry by post-processing the simulations allows us to consider much larger chemical networks that would otherwise be computationally impossible when evolving the chemistry in parallel. However, the chemistry is not coupled to the hydrodynamics, so, although {\sc 3d-pdr} calculates the gas temperature due to UV heating and cooling, it does not influence the gas evolution, and consequently the hydrodynamics and chemistry are not fully consistent. However, even for the warmer ( $\approx$ 100 K) gas, motions are dominated by turbulence rather than thermal broadening, so we do not expect large differences. For further discussion of the chemistry modelling, see \citet{2013ApJ...770...49O}. Even though the chemistry considered here is not time-dependent, the approach exhibits good agreement with \citet{2012MNRAS.421..116G}. While their chemistry network is considerably simpler than ours, the equilibrium time scales they find are the same order of magnitude of the times at which we evolve the chemistry to in this study, and are the same order of magnitude of the free fall timescale of the modelled cloud.

\subsection{Synthetic Emission Maps}\label{sec:sem}
We carry out synthetic observations for the 16 difference species in Table \ref{table:trans} and compare our theoretical results to observations. We study these species because they are commonly used tracers. We use the Leiden Atomic and Molecular Database (LAMBDA) \footnote{\href{http://home.strw.leidenuniv.nl/~moldata/}{http://home.strw.leidenuniv.nl/~moldata/}} for the reaction rates and cross sections for the different molecules. When performing these calculations, we use the collisional partners defined in the LAMBDA database files, mainly H$_2$, H and He, assuming most of the H$_2$ is para-H$_2$.

To compute the emission, we use the radiative transfer code {\sc radmc-3d} \footnote{\href{http://www.ita.uni-heidelberg.de/dullemond/software/radmc-3d/}{http://www.ita.uni-heidelberg.de/dullemond/software/radmc-3d/}} with the Large Velocity Gradient (LVG) approach \citep{2011MNRAS.412.1686S}, a radiative transfer method that does not assume local thermodynamic equilibrium (LTE). This method computes the molecular level populations given the density, velocity and temperature fields. We use the velocity information from the {\sc orion} calculation, while {\sc 3d-pdr} calculates the temperature and abundance information. Before performing the radiative transfer, we interpolate all data to a 256$^3$ resolution. We calculate the synthetic spectra for velocities within $\pm$ 3 km s$^{-1}$ of the line center. While {\sc 3d-pdr} computes the level populations for all levels in the LAMBDA data, we only analyze the emission from ground level transitions.

We use a constant microturbulence value of 0.1 km s$^{-1}$ to account for unresolved turbulence. We adopt a ``Doppler catching" parameter $d_c = 0.025$, which forces an interpolation of the velocity field between cells if there is a jump greater than 0.025 times the local linewidth. We use collisional excitation and line data from the Leiden atomic database for all 16 species \citep{2005A&A...432..369S}.

Since the main goal is to study the structure of the tracer species, we do not include the dust continuum in the emission calculation. We neglect heating and UV feedback due to embedded protostellar sources. However, since the cloud is forming low-mass stars, radiative feedback likely has a small impact on the emission. We convert the line emission into a brightness temperature $\rm{T_b}$ using the Rayleigh-Jeans approximation:
\begin{equation}
T_b = \frac{c^2 I_{\nu}}{2\nu_i^2 k_b}
\end{equation}
where $I_{\nu}$ is the specific intensity and $\nu_i$ is the frequency of the line transition.

\subsection{Statistical Analysis: Spectral Correlation Function}\label{sec:sa}

In this study, we use the Spectral Correlation Function (hereafter denoted as SCF), first introduced by \citet{1999ApJ...524..887R} to study the spectral structure of the emission cubes. We define a position-position-velocity (PPV) cube as spectral line data consisting of two spatial dimensions and one velocity dimension. Likewise, a position-position-position (PPP) cube is data consisting of the density information in all 3 spatial dimensions. The SCF is sensitive to the temperature and sonic Mach numbers and weakly sensitive to the the magnetic field stength \citep{2003ApJ...588..881P,2014ApJ...783...93Y}. Previous studies have only focused on the SCF of $^{13}$CO emission, in both simulated and observed molecular clouds. We calculate the SCF for all 16 species emission and density cubes to study how the SCF changes for the same cloud but different observational tracers. 

\begin{figure}[h!]
\begin{center}
\includegraphics[width=\columnwidth]{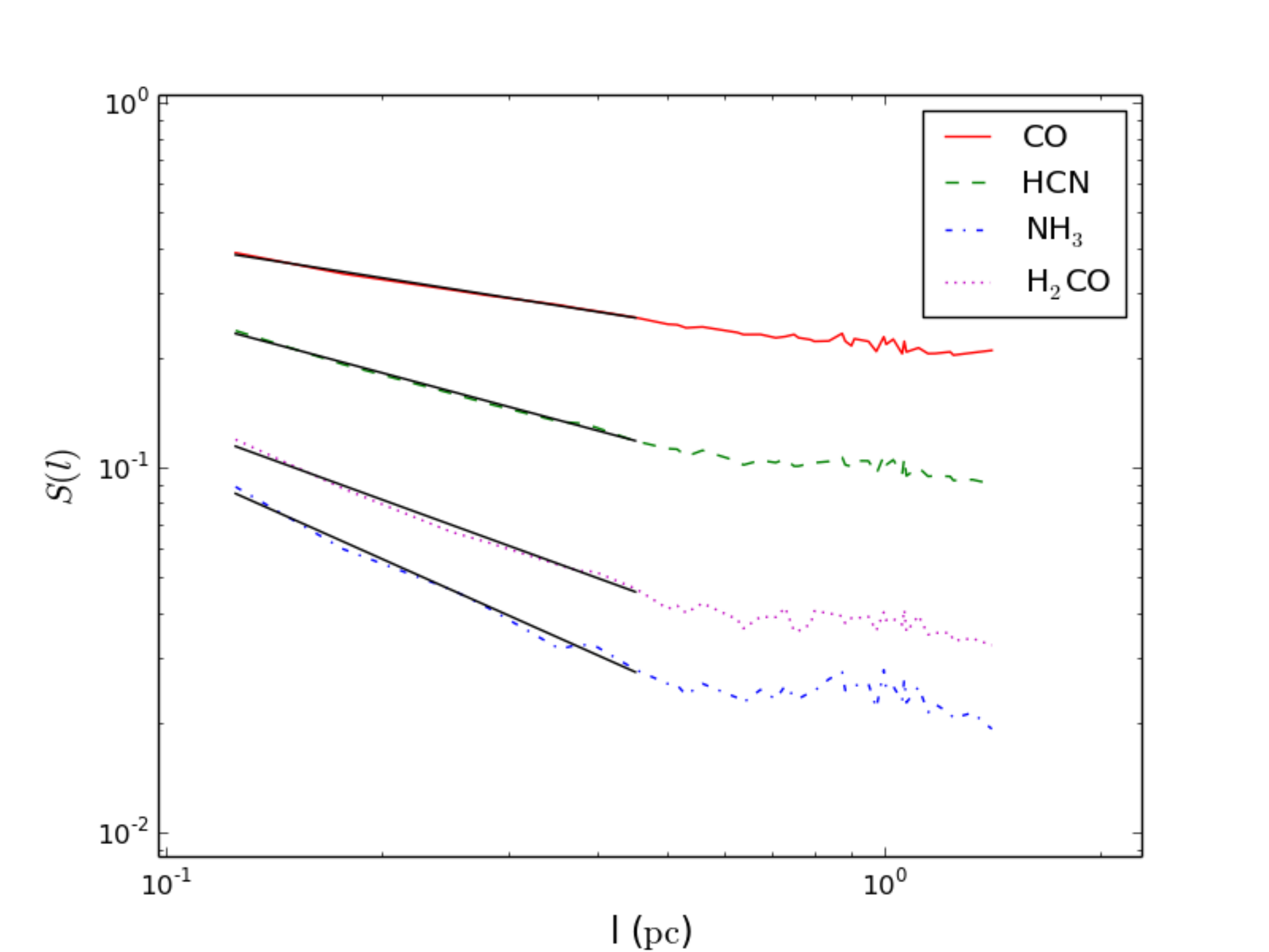}
\caption{\label{fig:scfs} Spectral correlation functions for four different species. The black lines indicate a power law fit.}
\end{center}
\end{figure}

There are several different functional forms for the SCF. We use the form given in \citet{2003ApJ...588..881P}, where the SCF is a function of length scale, $l$ and denoted by $S(l)$:
\begin{equation}
S(l) = \left < 1 - \left < \sqrt{\frac{\sum_v | O(\textbf{r}, v) - O(\textbf{r + l}, v)|^2}{\sum_v |O(\textbf{r}, v)|^2 + \sum_v |O(\textbf{r+l}, v)|^2}} \right >_{\textbf{r}} \right >_{|\textbf{l}| = l}
\end{equation}
Here $\textbf{r}$ is a two dimensional position on the image plane, $\textbf{l}$ is the offset vector with length $l$, and $O(\textbf{r}, v)$ is any PPV spectral data set. By definition $S(0) \equiv 1$. The SCF is defined to be bounded between 0 and 1, with 0 indicating no correlation. \citet{2003ApJ...588..881P} found that in driven turbulence simulations, the SCF can be analytically fit by a power law on small scales. They fit the SCF for $^{13}$CO (1-0) maps of several observed and simulated molecular clouds and demonstrated that the parameter correlations can be used as theoretical model tests. Likewise, we fit the SCF on small length scales using a power law:
\begin{equation}
S(l) = S_0 l^{\alpha}
\end{equation}
where $S_0$ is the value of the SCF at $l = 1$ pc. We fit the power laws in log space for length scales between 3$\times$ $l_{\rm min}$, where $l_{\rm min}$ corresponds to either the simulation resolution, or the beam size, and $l = $1 pc corresponding to half the maximum length scale. This was to remove beam size effects from the SCF power law fit. Figure ~\ref{fig:scfs} shows the SCF for four different tracers. The tracers follow a tight power law behavior for small values of $l$, as illustrated by the black lines fits. However, Figure ~\ref{fig:scfs} also shows that at some length scale, the SCF function appears to flatten and become noisy. 

When comparing SCFs for different spectral maps, it is useful to define a quantitative scalar value that describes how similar two SCFs are to one another. We use the distance metric between two SCFs as defined by \citet{2014ApJ...783...93Y}:
\begin{equation}
d_{SCF} = \sqrt{\sum_l [S_1(l) - S_2(l)]^2}
\end{equation}
Using this distance metric provides a quantitative measure of how similar the spectral structures are between different species, resolutions, and sight lines. \citet{2014ApJ...783...93Y} showed that this metric is sensitive to global hydrodynamic parameters, although not to the magnetic field strength. Since the distance metric is a 1D statistic we can't compute a chi-squared value, i.e. a probability measure of uncertainty. In \S\ref{sec:va} we will describe an example of using the distance metric to quantitatively comparing SCFs.

\section{Results}
\subsection{Power Law Range}\label{sec:pl}
Figure \ref{fig:scfs} shows that the SCFs of species which trace dense gas transition away from power law behavior at larger scales and shows that the power law description holds only up to some scale $l \ll L$ where $L$ is the cloud size. At larger length scales, the SCFs start to flatten for all species. For example, the CO SCF is a continuous power-law up to nearly 2 pc, while the SCF for NH$_3$ starts to flatten around 0.4 pc. This flattening indicates where the gas is not dense enough to thermally emit. CO remains correlated at large scales throughout the cloud, with a high surface filling fraction of $f_s = 0.82$, and samples both diffuse and dense structures. CO has an low volume filling fraction, $f_v = 0.07$ since it is still contained mostly in dense environments with emission coming from diffuse regions because of its substantially smaller critical density. The emission from high density regions is constrained by its high optical depth. In our study, the surface filling fraction, $f_s$, is the fraction of the area containing the brightest 95\% of the emission. The volume filling fraction, $f_v$, is defined from the 3D abundances, being the fraction of the volume containing the top 95\% of the mass. However, NH$_3$ becomes quickly uncorrelated, because it traces only compact emission, i.e. dense cores, illustrated its much smaller volume filling fraction, $f_v \sim 0.005$ and small surface filling fraction $f_s \sim 0.25$. 

\begin{figure}[h!]
\begin{center}
\includegraphics[width=\columnwidth]{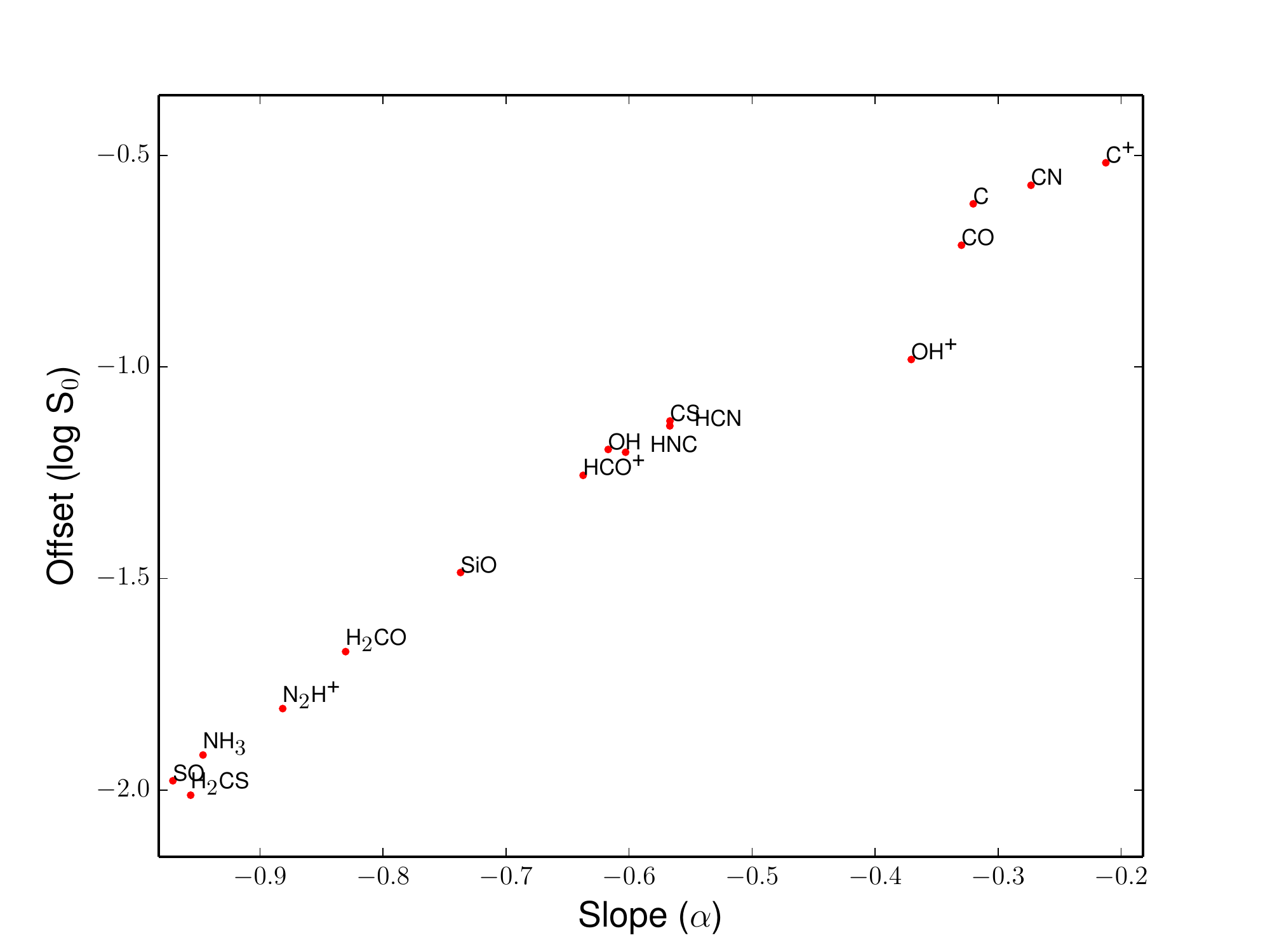}
\caption{\label{fig:so} Slope, $\alpha$, versus offset, $S_0$, for the SCFs calculated for 16 PPV emission maps. The same trend holds for the density data, so we only show the emission parameters.}
\end{center}
\end{figure}

Figure ~\ref{fig:so} displays the SCF slope as a function of offset, $S_0$ for 16 synthetic emission cubes, where $S_0$ is the value of the SCF at 1 pc. While we define the SCF in terms of a slope and an offset, we show in Figure ~\ref{fig:so} that these two parameters are tightly correlated. Therefore, henceforth, we only discuss the slope of the SCF. \citet{2003ApJ...588..881P} also found a strong correlation between the slope and the offset for simulated clouds.

\subsection{Viewing Angle}\label{sec:va}
A statistic describing cloud structure is only meaningful if it does not vary strongly as a function of cloud viewing angle. Here we verify that this is the case. 

\begin{figure}[h!]
\begin{center}
\includegraphics[width=\columnwidth]{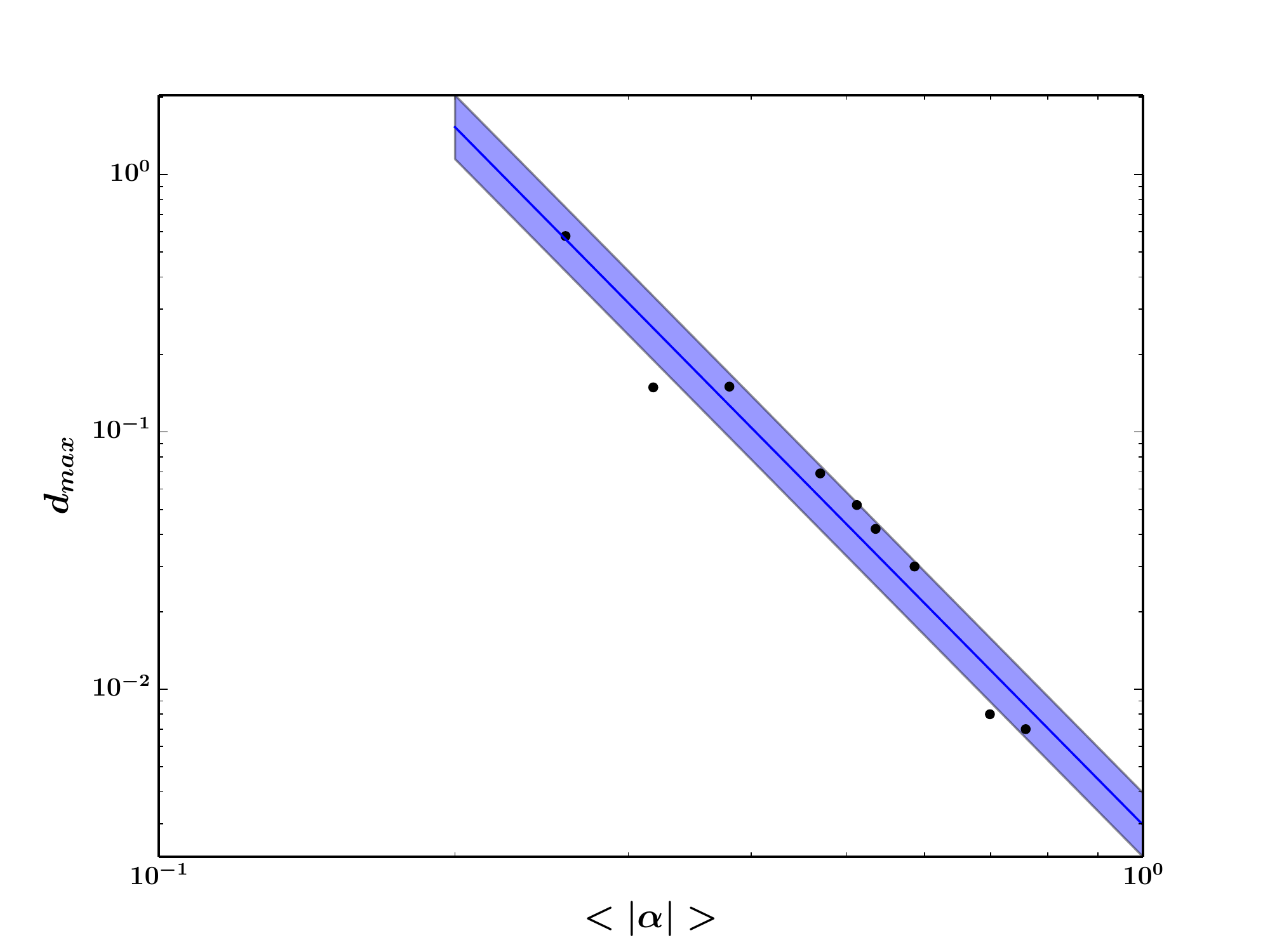}
\caption{
\label{fig:dalph} Maximum SCF distance, $d_{max}$, as a function of the average of the SCF slope magnitude, $<|\alpha|>$, for different tracers (black dots). The blue line indicates a power law fit to the points with a slope of -3.2. The shaded region shows $\pm$ 1 $\sigma$ from the fit.}
\end{center}
\end{figure}

We calculate the emission for 9 different lines of sights through the cloud in 45 degree increments about the x-axis for all 16 tracers. We find that the SCF slopes change by only $\approx$ 10\% over all viewing angles. This variation, though small, is useful as a benchmark to define when two tracers are truly similar. To define an effective uncertainty, we calculate the maximum distance between any two SCFs for all lines of sight. Figure ~\ref{fig:dalph} shows this maximum distance for 6 different species. We find that there is a power law relation  between the average slope, $<|\alpha|>$, and the maximum distance, $d_{max}$. This relation now provides a quantitative way of identifying homologous tracers. We define two tracers as ``complementary" if their SCF distance for a given line of sight is less than the maximum $d_{\rm{max}}$ of either tracer. Therefore, species $A$ and $B$ are complementary if
\begin{equation}
d_{\rm SCF}(B \Leftrightarrow A) \leq {\rm max}(d_{\rm max}(A), d_{\rm max}(B))
\end{equation}
Figure \ref{fig:dalph} also suggests that diffuse species are more likely to have complements because they have shallower slopes. We can see this by comparing the SCFs of CO and NH$_3$. If we fix the CO offset and alter the slope by the maximum $\pm$ 10\% error, the total rms difference is geometrically $d_{max}$. Since the SCF slope is shallower for CO, the magnitude of the SCF remains higher. In contrast, if the NH$_3$ SCF is altered in a similar manner, the magnitude of the SCF becomes smaller much faster as a result of the steeper power law. Therefore, the rms difference between the SCFs is much smaller than what it would be for CO. As such, there is an expectation that $d_{max}$ should decrease with increasing slope.

Figure \ref{fig:dalphC} shows the SCF slopes for all the species in this study with their line of sight scatter. The species are seemingly separated into three distinct groups, though only two of the groups show a large jump in slope. Based on these groupings, we will refer to {\it diffuse} tracers as those with $\alpha \sim -0.3$, {\it intermediate} tracers as those with $\alpha \sim -0.5$, and {\it dense} tracers as those with $\alpha \sim -0.75$. CO and C both have the same SCF slope within 1 $\sigma$ of each other and within 2 $\sigma$ of all the other diffuse tracers except C$^+$, where $\sigma$ is the line of sight slope error. The larger errors for C$^+$ seem to correspond to different diffuse structures being superimposed at different sight lines giving somewhat different structures. Note that while the species whose SCFs are within the line of sight error limits are homologous (meaning they trace similar density regimes), they are only complementary if they also satisfy the distance metric criteria (Equation 5).

The simulated molecular cloud that we study has no magnetic fields. While \citet{2014ApJ...783...93Y} find that the SCF only weakly depends on the magnetic field, the presence of a strong magnetic field could create asymmetries in the gas distribution. Large asymmetries could in turn increase the SCF dependence on the viewing angle.
\subsection{Chemical Species}\label{sec:chemspec}
\begin{deluxetable*}{ccccccc}
	\tablecolumns{7}
	\tablewidth{\textwidth}
	\tablecaption{Chemical Species and Properties \label{table:trans}}
	\tablehead{
	\colhead{Name} & \colhead{Transition} & \colhead{Frequency (GHz)} & \colhead{$n_c$ (cm$^{-3}$)} & \colhead{$f_s$} & \colhead{$f_v$} & \colhead{Notes}}
	\startdata 
	C & J=1-0 & 492.160651 & 820 & 0.87 & 0.45 & \\
	C$^{+}$ & J=$\frac{3}{2}-\frac{1}{2}$ & 1900.5369 & 7700 & 0.87 & 0.48 & \\
	CN & N=1-0, J=$\frac{1}{2}-\frac{1}{2}$ & 113.1686723 & 1.3$\times$10$^{6}$ & 0.87 & 0.14 & \\
	CO & J=1-0 & 115.2712018 & 2180 & 0.82 & 0.07 & \\
	CS & J=1-0 & 48.9909549 & 5$\times$10$^{4}$ & 0.73 & 0.05& \\
	HCN & J=1-0 & 88.6316023 & 1$\times$10$^{6}$ & 0.65 &  0.40 & No hfs\\
	HCO$^{+}$ & J=1-0 & 89.188523 & 1.6$\times$ 10$^{5}$ & 0.62 & 0.14 & \\
	HNC & J=1-0 & 90.663568 & 2.7$\times$10$^{5}$ & 0.21 & 0.01 & \\
	OH & J=$\frac{3}{2}$,F = 1,P = +/- & 1.66 & 2.60 & 0.22 & 0.15 & No hfs \\
	OH$^{+}$ & N=1-0 & 909.15880 & 4270 & 0.63 & 0.36 & \\
	H$_2$CO & J=1-0 & 72.837948 & 1.5$\times$10$^{5}$ & 0.41 & 0.03 & p-H$_2$CO, hfs\\
	H$_2$CS & J=1-0 & 34.3543 & 8400 & 0.24 & 0.006 & p-H$_2$CS\\
	NH$_3$ & (J,K)=(1,1) & 23.6944955 & 1990 & 0.25 & 0.005 & p-NH$_3$, hfs\\
	N$_2$H$^{+}$ & J=1-0 & 93.17370 & 1.4$\times$10$^{5}$  & 0.21 & 0.005 & No hfs\\
	SiO & J=1-0 & 43.42376 & 3.8$\times$10$^{4}$ & 0.51 & 0.005 &\\
	SO & J=1-0,N=0-1 & 30.00158 & 7.7$\times$10$^{4}$ & 0.28 & 0.005 & 
	\enddata
	\tablecomments{Species name, transition, critical density, surface filling fraction and volume filling fraction for all the species in which we do radiative transfer post processing. We define the surface filling fraction as the fraction of the area that constitutes 95\% of the total intensity. The volume filling fraction is the fraction of the volume that constitutes 95\% of the total mass. Hyperfine splitting is denoted by hfs. Species that have hfs defined in the LAMBDA database files but are not included in the radiative transfer modelling are denoted by ``No hfs", while those that have hfs that is included in the radiative transfer modelling are indicated by ``hfs". Critical densities were calculated for gas at T = 10 K.} 
\end{deluxetable*}
We use 16 common astrophysical tracers to investigate how the SCF depends on species. We generate synthetic emission maps for the lowest energy state transitions shown in Table ~\ref{table:trans}. Figures ~\ref{fig:xvsy} and  ~\ref{fig:xvsv} show integrated emission maps in position-position (PP) and position-velocity (PV) space, respectively. The various tracers span different ranges of position and velocity space depending on their abundance and excitation. For example, astronomers commonly use N$_2$H$^{+}$ to trace dense gas, so it is unsurprising that N$^{2}$H$^{+}$ exhibits very compact emission in both figures.  The velocity plots indicate structure that may be hidden by projection. Dense cores stand out in both maps. 

We find that the filling fractions in PP and PV are similar: tracers with compact spatial emission are also compact in velocity space. This is essentially Larson's relation \citep{1981MNRAS.194..809L}, which states that small structures should have small velocity extents. CO has a high surface filling factor ($f_s \sim 0.82$) and traces both high and low density regions, with a relatively high abundance across the entire spatial region, exhibited by its spatial emission distribution. CO is chemically connected to the high density tracer H$_2$CO through a number of reactions. H$_2$CO is photodissociated to form CO when the density becomes low enough that it is no longer shielded from the external UV field. CO can get turned into H$_2$CO through intermediaries, such as HCO$^+$ in both gas phase reactions and dust grain chemistry. As expected then, the H$_2$CO emission has a much smaller surface filling fraction of $f_s \sim 0.41$ and is resides mostly in dense environments, requiring a relatively high density to be excited (n$_c \sim 10^5$ cm$^{-3}$).

Optical depth indicates the degree of transparency. Tracers that are optically thin, such as NH$_3$, have emission that reflects the underlying density structure more accurately. Figure \ref{fig:tau} shows the optical depth of each line transition at the line center calculated by {\sc radmc-3d}. The figure shows that C, CO, CN and C+ all have high optical depths. This is a result of having a lower critical density with a relatively high abundance. The intermediate density tracers, CS, SiO, HCN, HNC and HCO+ are similar, with the gas only being optically thick towards the highest density regions. High density tracers, NH$_3$, H$_2$CO, H$_2$CS and N$_2$H$^+$ remain optically thin throughout almost all of the entire volume. NH$_3$ is an exception because of its fairly low critical density, allowing it to be excited down to lower densities, though still relatively optically thin except in the densest regions of the filaments. OH$^+$ suffers from a very low abundance, and it is not easy to excite, so its emission appears very diffuse.

\begin{figure}[h!]
\begin{center}
\begin{tabular}{c}
\includegraphics[width=\columnwidth]{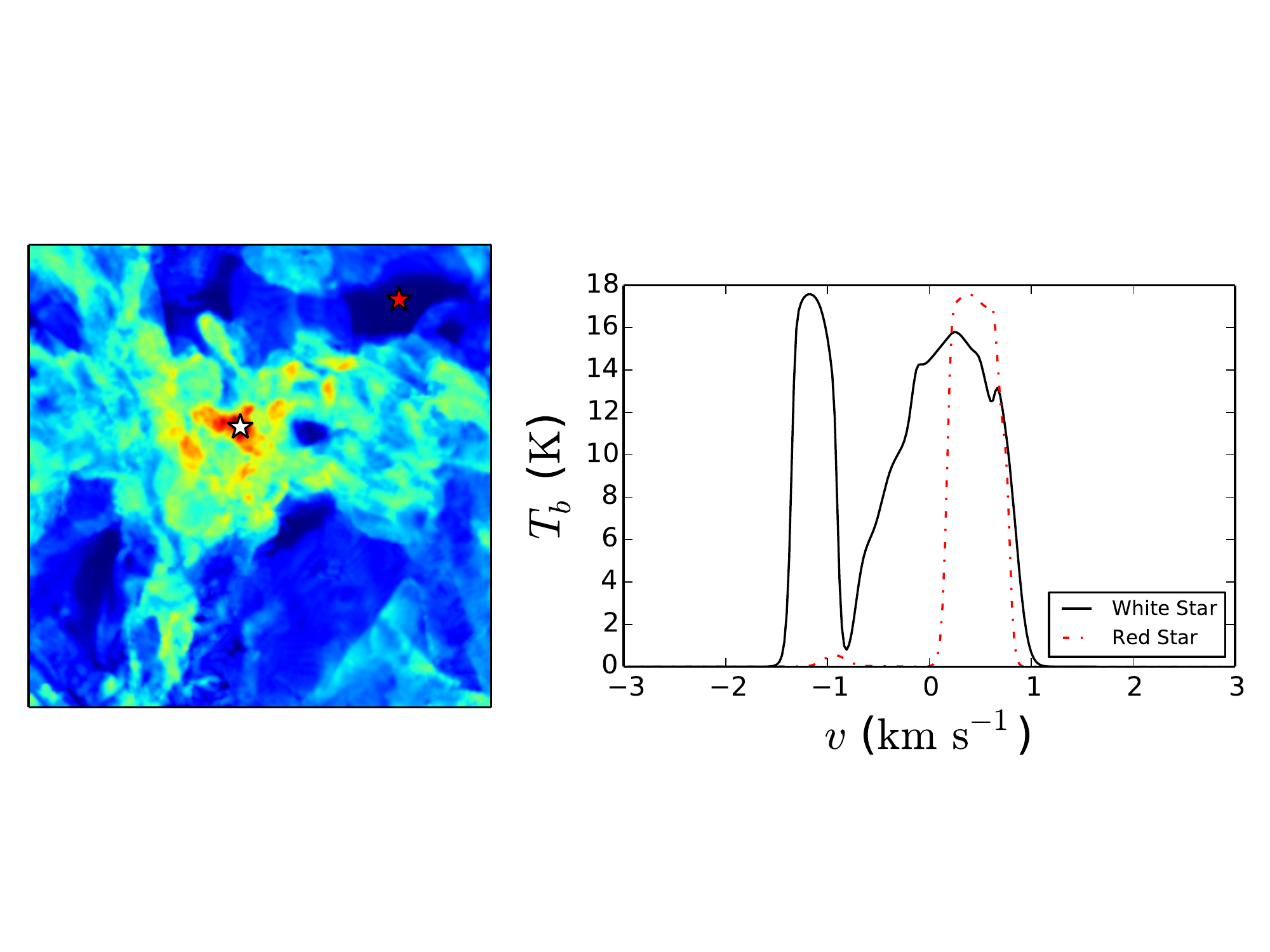} \\
\includegraphics[width=\columnwidth]{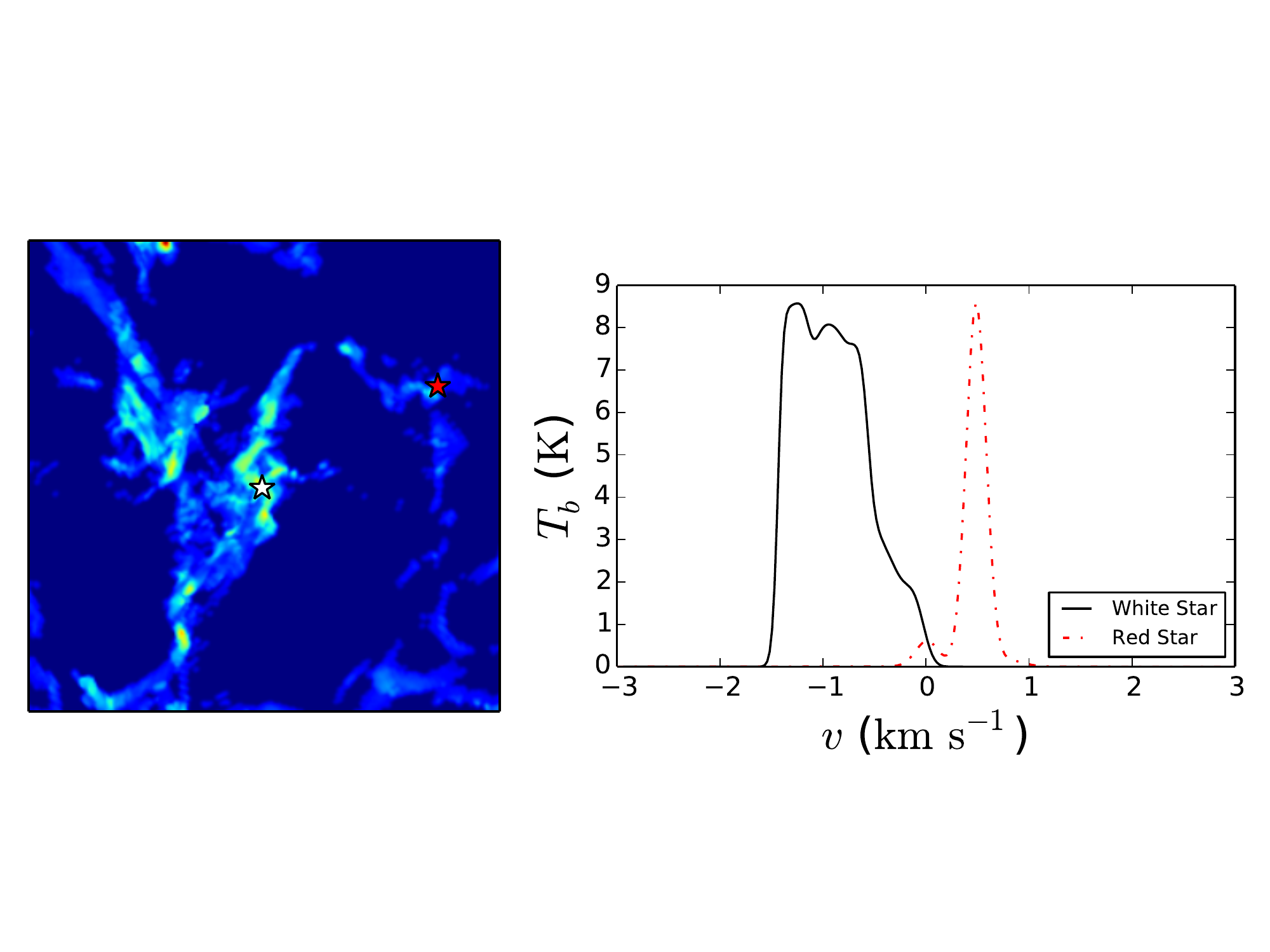}
\end{tabular}
\caption{\label{fig:2plot} Integrated emission maps (left) and line-of-sight velocity spectra (right) for CO (top) and NH$_3$ (bottom). The white and red star locations represent emission from compact regions and diffuse regions, respectively.}
\end{center}
\end{figure}

Figure ~\ref{fig:2plot} displays two different integrated maps with spectra at two different points. It shows that the CO emission is mostly dominated by gas motions. The red star spectrum shows a single feature with a width of $\sim$ 0.7 km s$^{-1}$, which is consistent with the characteristic turbulence velocity. The CO white star spectrum is broad in part because it is a superposition of multiple density features along that particular line of sight. Similarly, the NH$_3$ white star spectrum shows a superposition of several density features at different velocities along the line of sight. However, the NH$_3$ red star spectrum, which is along a rather diffuse line of sight, shows a much narrower feature with a width of $~ 0.3$ km s$^{-1}$. The average sound speed of a cold molecular cloud is approximately 0.2 km s$^{-1}$, indicating that the NH$_3$ in the red star region is undergoing purely thermal motion rather than dominated by turbulence.

\subsection{Resolution}\label{sec:res}
\begin{figure}[h!]
\begin{center}
\includegraphics[width=\columnwidth]{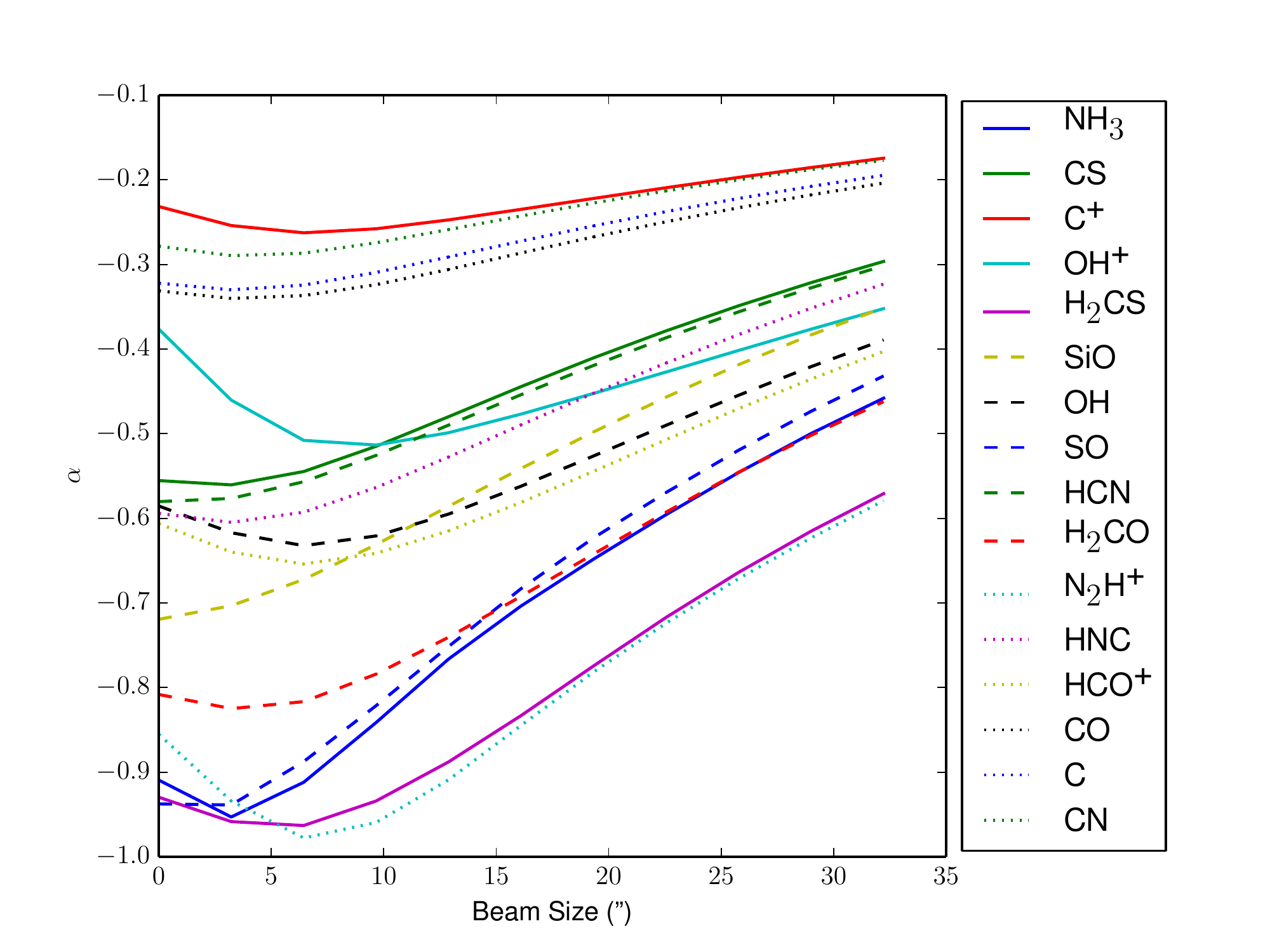}
\caption{\label{fig:sn} SCF slope, $\alpha$, versus beam resolution.  We calculate the angular beam size by placing the model cloud at a distance of 1 kpc. The clouds are then convolved with a Gaussian of the given beam size.
}
\end{center}
\end{figure}

Beam resolution is one of the most important factors in observations and it impacts the apparent gas structure and mean optical depth. A larger beam averages out the emission within the beam size, lowering the overall optical depth. We convolve the emission maps with a Gaussian beam to simulate a realistic resolution observation. We place the simulated cloud at a distance of 1 kpc to establish the angular size. Larger beams significantly blend the dense cores and diffuse gas structure. Interestingly, we find that some species' spectral structures are artificially similar at one resolution and then deviate significantly at some lower resolution. To quantify this, at each resolution we calculate the SCF for each species and then compare their distance metrics. Figure ~\ref{fig:sn} shows the evolution of the SCF slopes for all 16 species as a function of spatial resolution. The overlapping regions indicate where tracers at a particular spatial resolution have similar emission distributions. Note that this only impacts the emission. Therefore, overlapping regions do not indicate similar density distributions, unless the SCF slopes are similar at good (near simulation) resolutions. Thus, Figure ~\ref{fig:sn} quantifies which tracers are statistically similar and useful for studying particular densities and size scales.

At the highest resolution (i.e. no smoothing),  C, CO, CN and C$^{+}$ are all very similar tracers since they all trace diffuse gas. The positive correlation between C and $^{12}$CO has been observationally studied for years \citep[e.g.][]{1999ApJ...512..768P, 1999ApJ...527L..59I, 2002ApJS..139..467I, 2005ApJ...625..194K, 2013ApJ...774L..20S}, although historically C was theoretically predicted to only exist in a PDR surface layer \citep{1999RvMP...71..173H}. \citet{2004MNRAS.351..147P} predicted that C should be more prevalent in clouds than previously predicted by 1D models due to a combination of non-equilibrium processes and clumpy cloud morphology. \citet{2014arXiv1403.3530G} and \citet{2014MNRAS.tmpL..37O} both qualitatively demonstrated the similarity between the C and CO distributions in 3D PDR calculations. Our SCF comparison demonstrates that CO and C spectral cubes are {\it quantitatively} similar.

At higher densities, species such as SO and NH$_3$ also appear very similar. As expected, HCN and HNC structures appear nearly identical in spectral space. At lower spatial resolution, several species intersect: NH$_3$, N$_2$H$^{+}$, SO and H$_2$CO. This occurs as larger beams increasingly blend dense, compact emission.

\begin{figure}[h!]
\begin{center}
\begin{tabular}{c}
\includegraphics[width=\columnwidth]{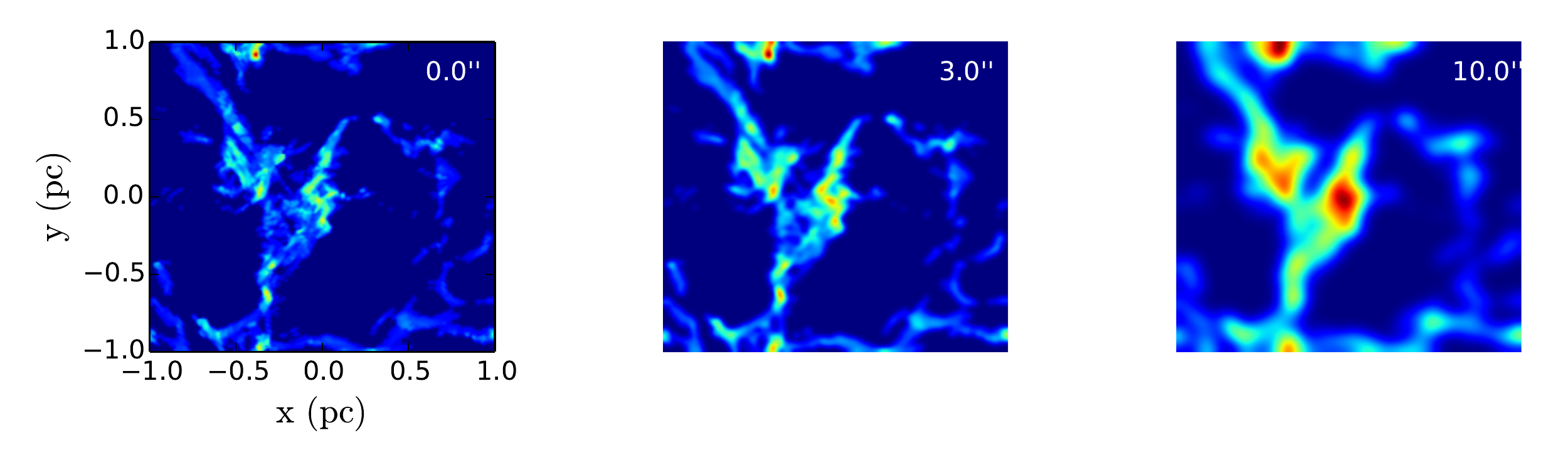} \\
\includegraphics[width=\columnwidth]{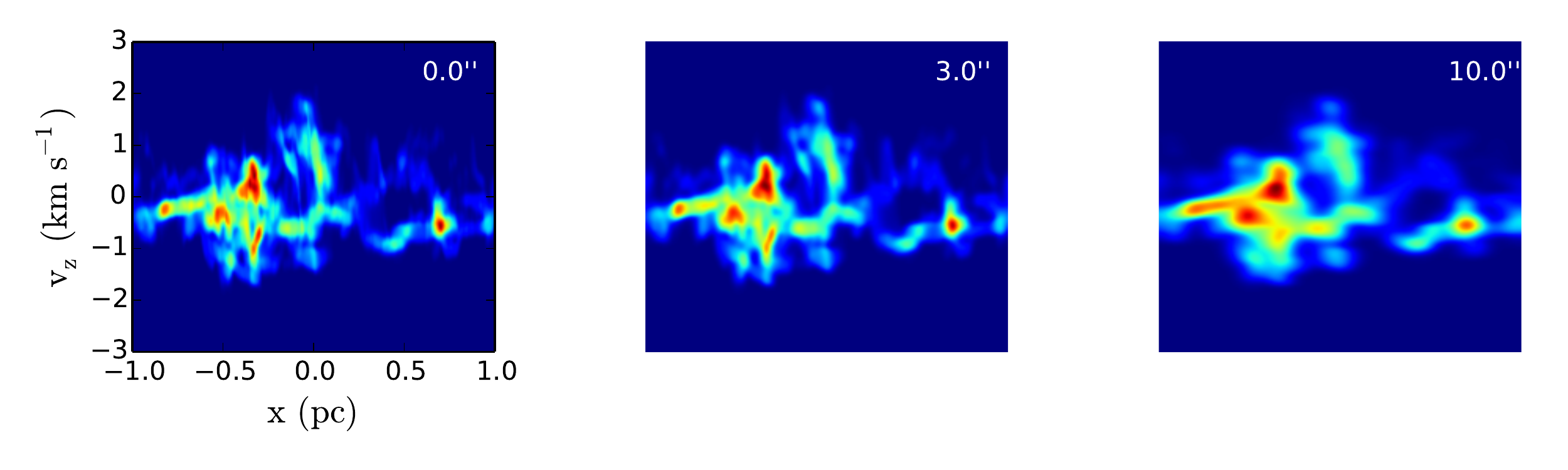}
\end{tabular}
\caption{\label{fig:deres} NH$_3$ integrated emission maps for PP (top) and PV (bottom) at different spatial resolutions for a distance of 1 kpc. The beam size appears in the top right.}
\end{center}
\end{figure}

\begin{figure*}[h!]
\begin{center}
\hspace{-0.1\columnwidth}
\begin{tabular}{c c}
\includegraphics[width=0.55\textwidth]{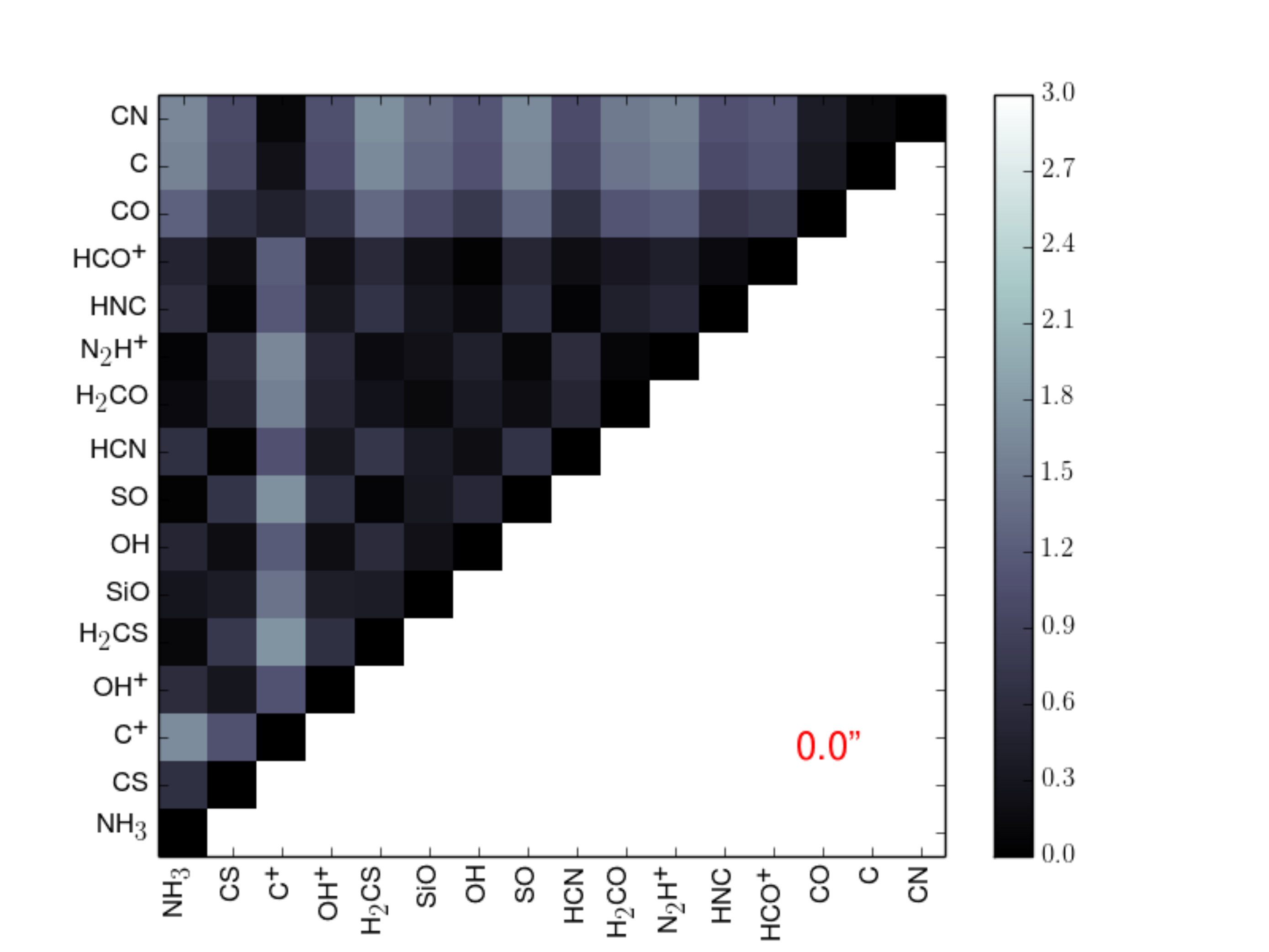} & \includegraphics[width=0.55\textwidth]{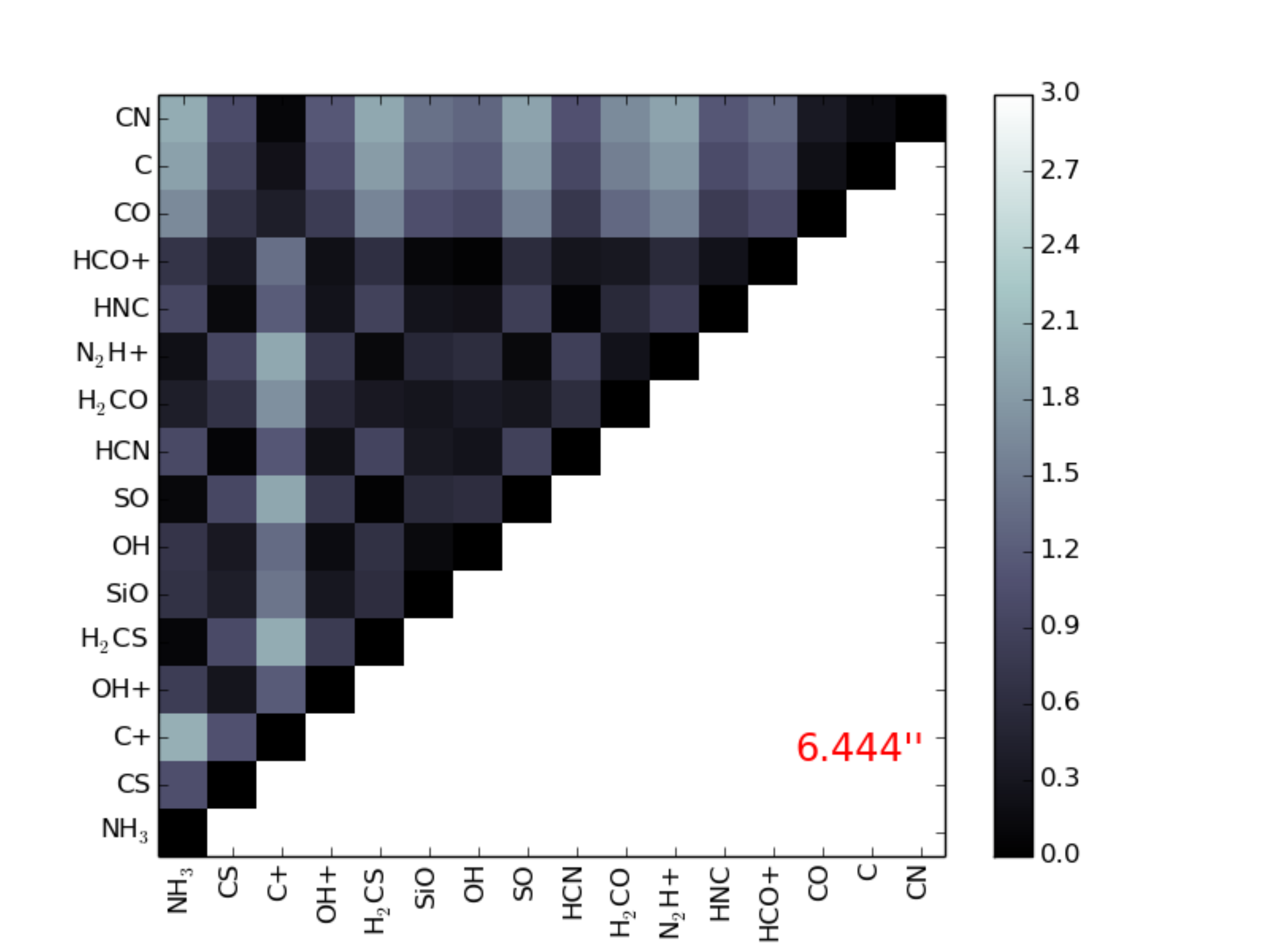} \\
\includegraphics[width=0.55\textwidth]{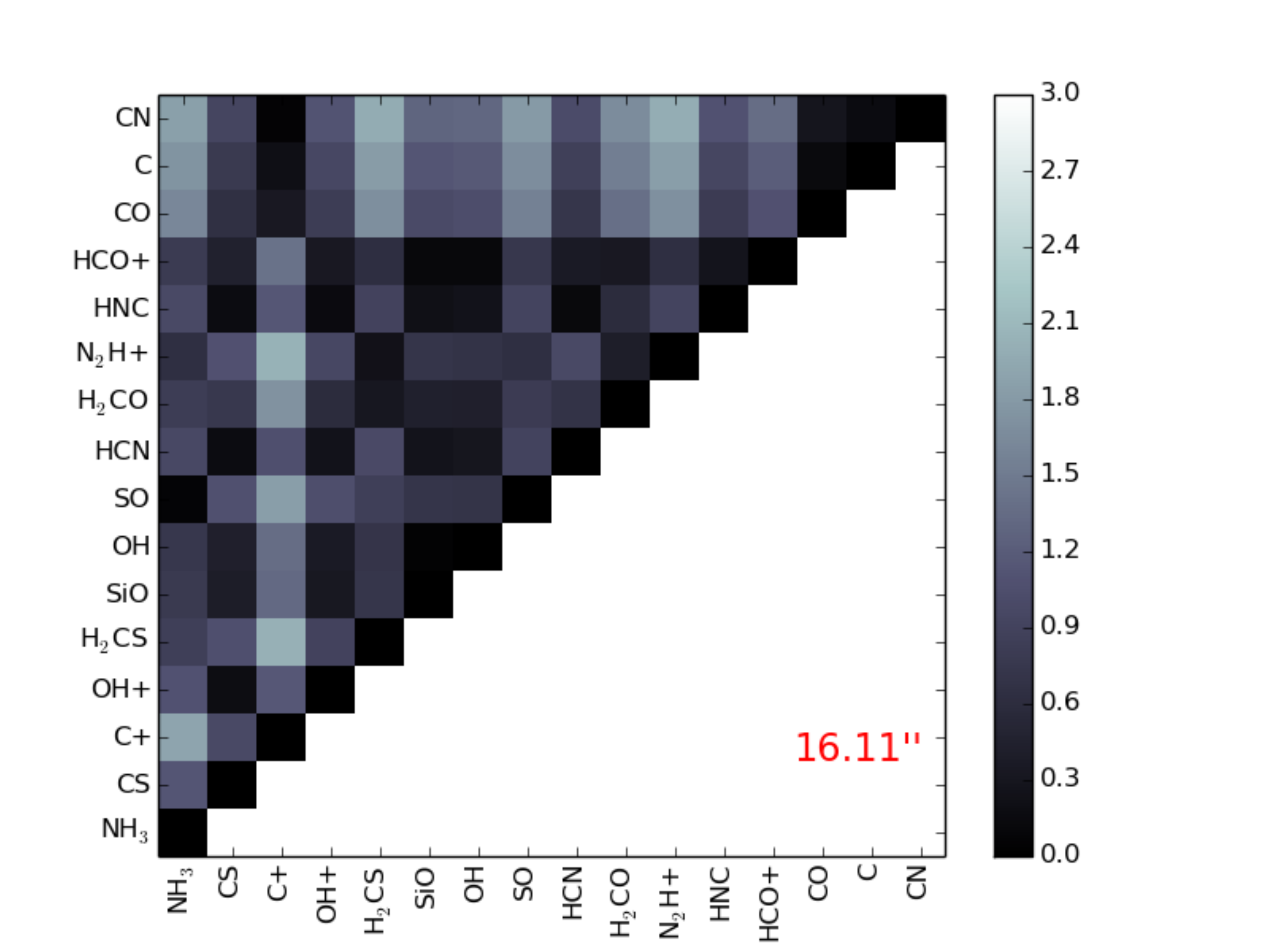} & \includegraphics[width=0.55\textwidth]{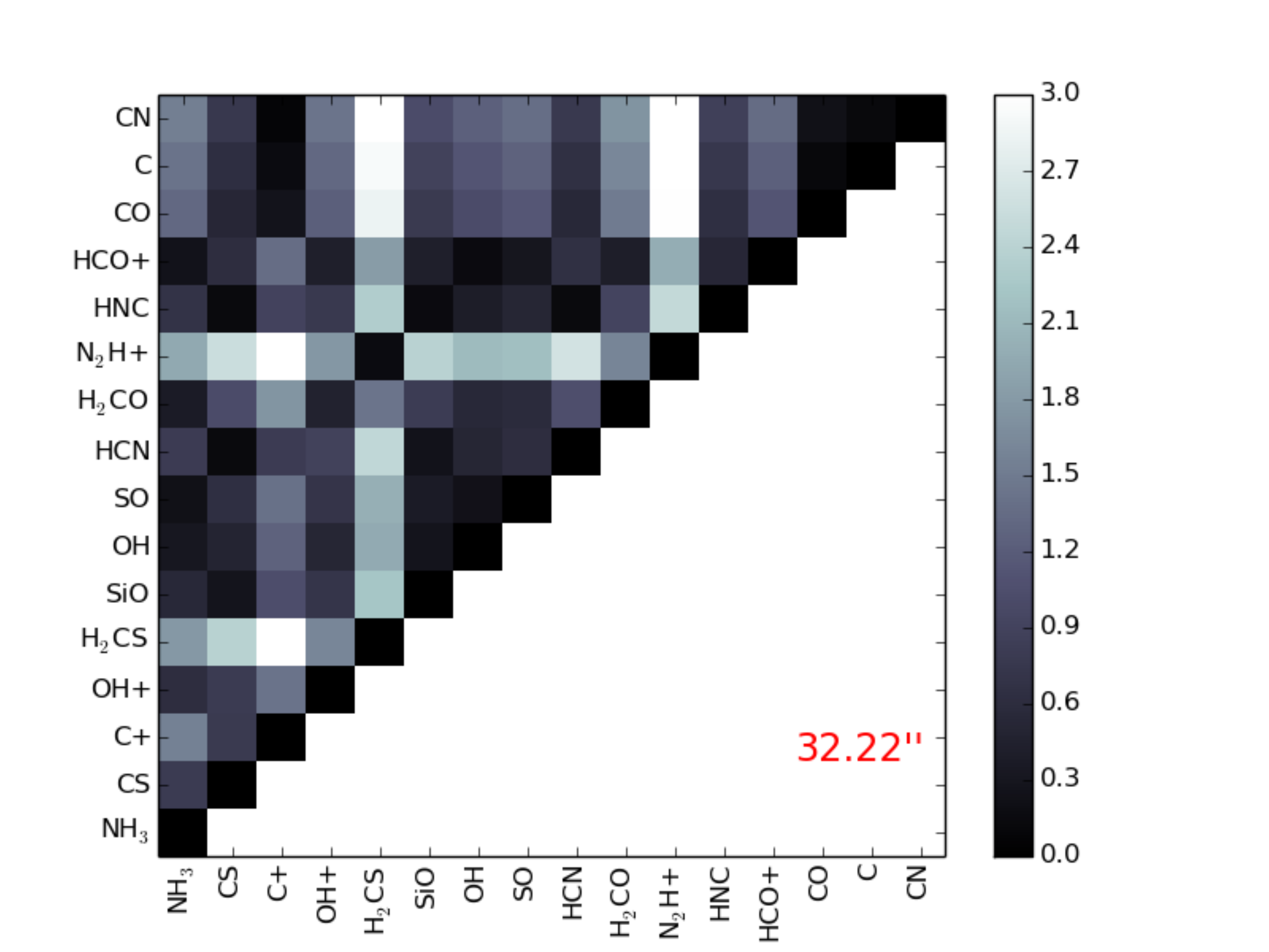}
\end{tabular}
\caption{\label{fig:dE} Distant metrics for all 16 species, where darker colors indicate more similar SCFs. We calculate the resolution by setting the cloud at a distance of 1 kpc  for the beam size shown in the bottom right.
}
\end{center}
\end{figure*}

We note that some species have a non-monotonic dependence on resolution where the SCF slope decreases until $\sim$ 15-20'' and then increases again. The decrease in the slope is due to overlapping Gaussian structures creating artificial cores. Figure \ref{fig:deres} shows the emission spectral structure evolution with beam size visually for NH$_3$ emission in both PP and PV space. Figures \ref{fig:dE} and \ref{fig:dOE} illustrate these trends as a function of beam size using the distance metric. The change in the slope occurs because species with significant compact emission (those which are generally optically thin) have their emission smoothed on larger scales, making their emission more extended and thus appear similar to other more optically thick species. Smoothing also somewhat affects the more diffuse tracers C, CO, CN and C+, which experience blending on larger scales. This can give the appearance of a more core-like structure where these smoothed regions overlap. At lower density, the emission smooths into am even more diffuse looking component.

\subsection{Comparison of Density and Emission}\label{sec:comparisonDE}
\begin{figure*}[h!]
\begin{center}
\hspace{-0.1\columnwidth}
\begin{tabular}{c c}
\includegraphics[width=0.55\textwidth]{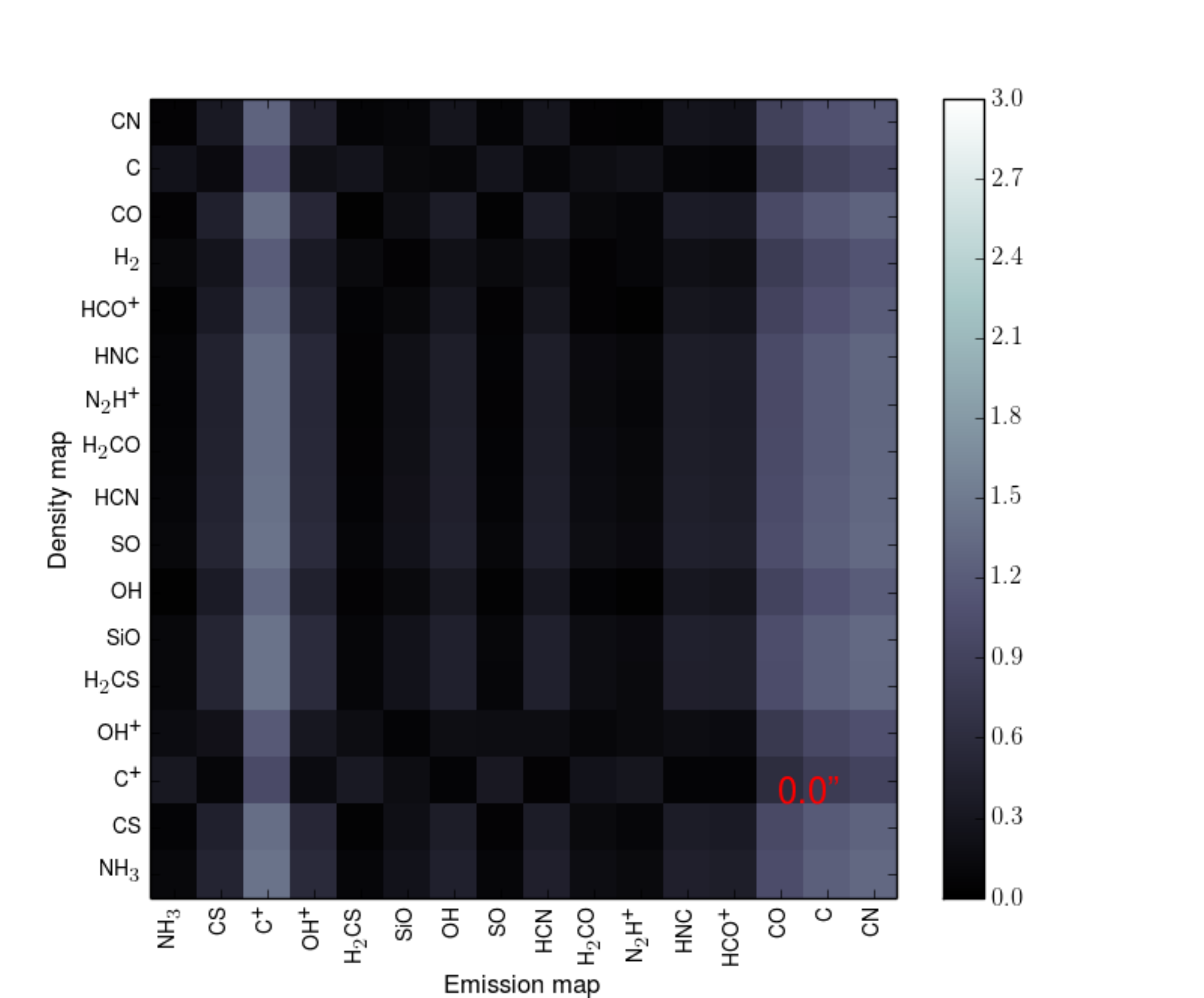} & \includegraphics[width=0.55\textwidth]{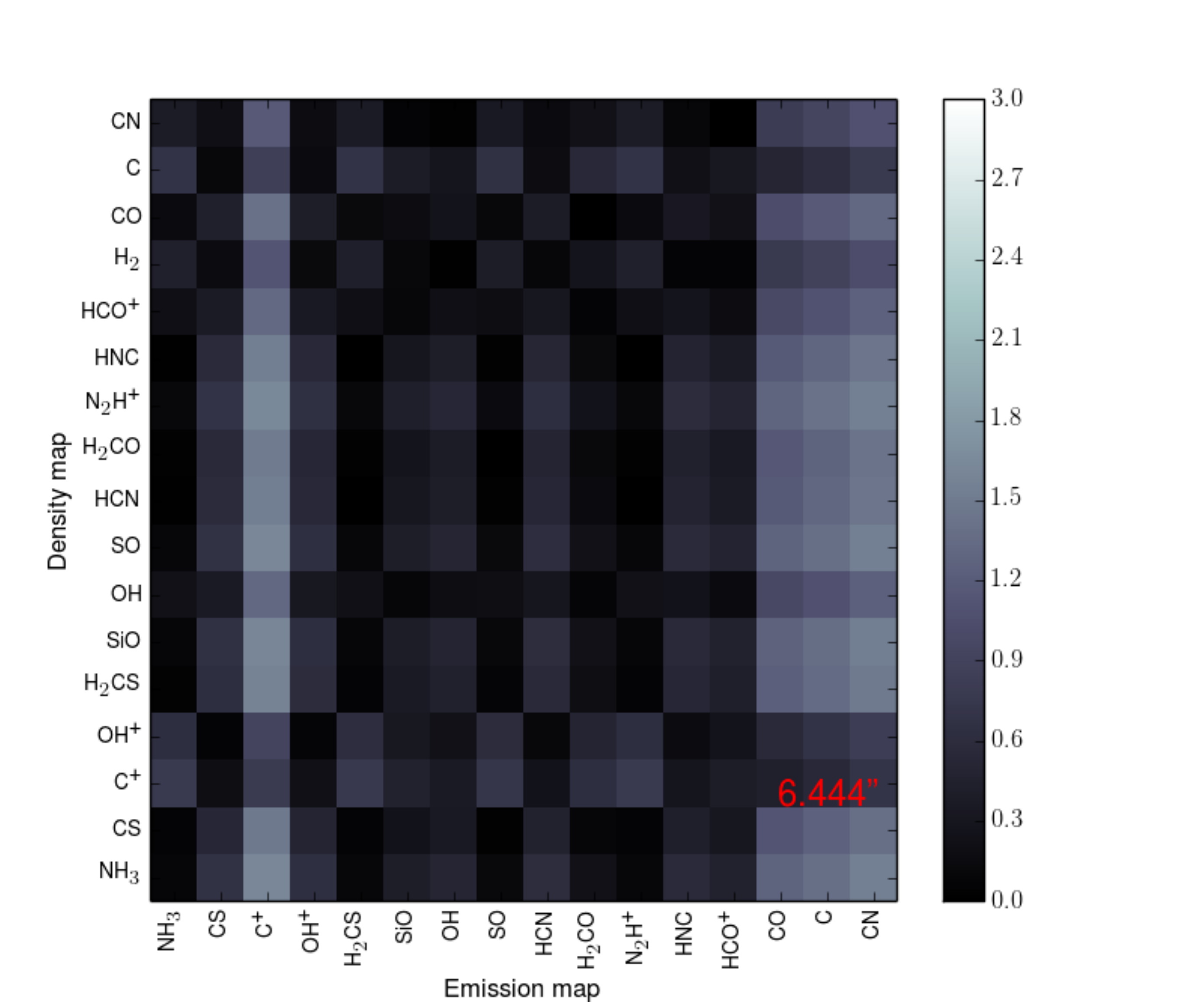} \\
\includegraphics[width=0.55\textwidth]{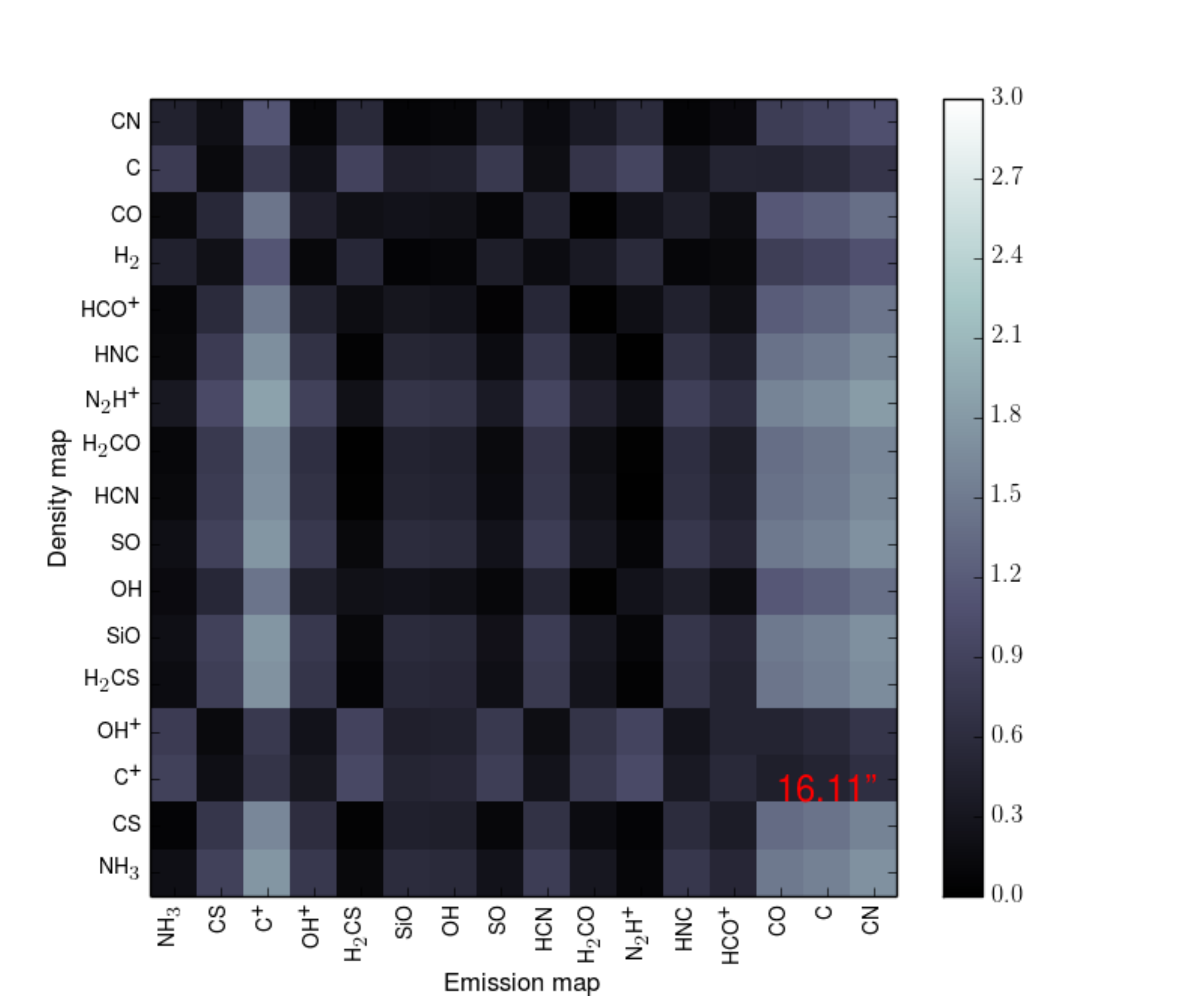} & \includegraphics[width=0.55\textwidth]{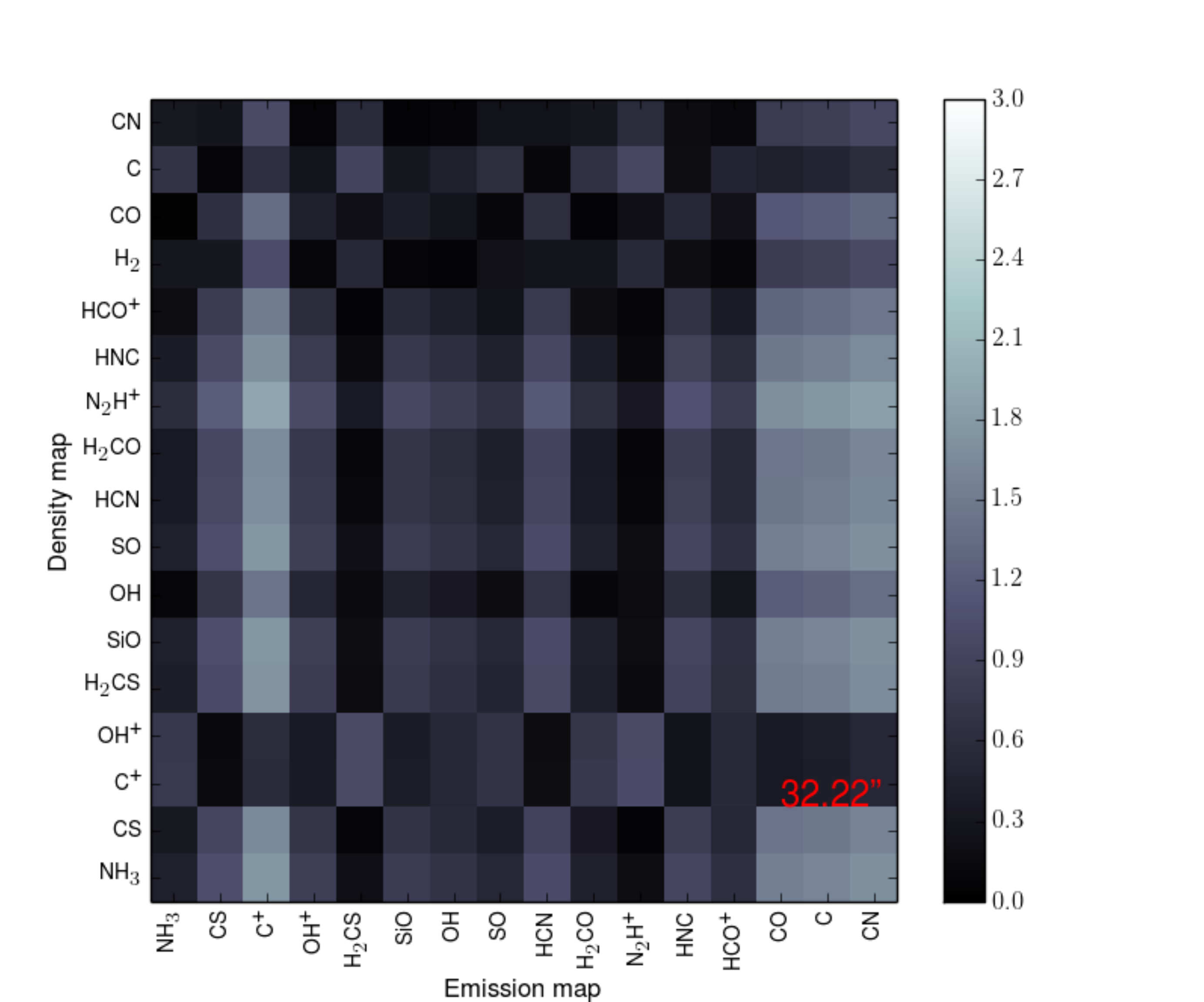}
\end{tabular}
\caption{\label{fig:dOE} Distance between the density PPV cube SCF (vertical) and the intensity PPV cube SCF (horizontal) for each species pair. The beam resolution for the intensity PPV cube as placed at a distance of 1 kpc, appears at the bottom right. The PPV$_{\rm \rho}$ cube is not blurred with the beam.
}
\end{center}
\end{figure*}

In \S3.3 we demonstrate that many tracers produce similar SCFs. However, similarity between the SCFs of two emission maps does not guarantee that the underlying densities are also similar. In this section, we compare the SCFs of the emission and density data directly. We obtain a PPV cube based on the gas density (hereafter denoted by PPV$_{\rho}$) by taking the simulated density and velocity cubes and constructing a PPV cube where each spectral bin contains the number density (calculated as $N_i = N_{\rm H}\cdot n_i$) within a given velocity bin. Here, ``V" is the velocity vector projected along a given line of sight. For the PPV$_\rho$ cube, the (i,j,k)$^{\rm th}$ voxel contains the total mass along some line of sight through the point (x$_i$, y$_i$) within a particular line of sight velocity range $\Delta$v$_k$. We compare the emission and density SCFs in Figure ~\ref{fig:dOE}. This figure shows the distance between the emission SCFs and PPV$_{\rho}$ SCFs for each species. For C, CO, CN and C+, PPV$_{\rho}$ does not match the emission structure well. In fact, as the spatial resolution of the PPV cube becomes coarser, agreement between density and emission worsens. High opacity tracers exhibit poor correspondence between emission and density. 

\begin{figure*}[h!]
\begin{center}
\includegraphics[width=0.8\textwidth]{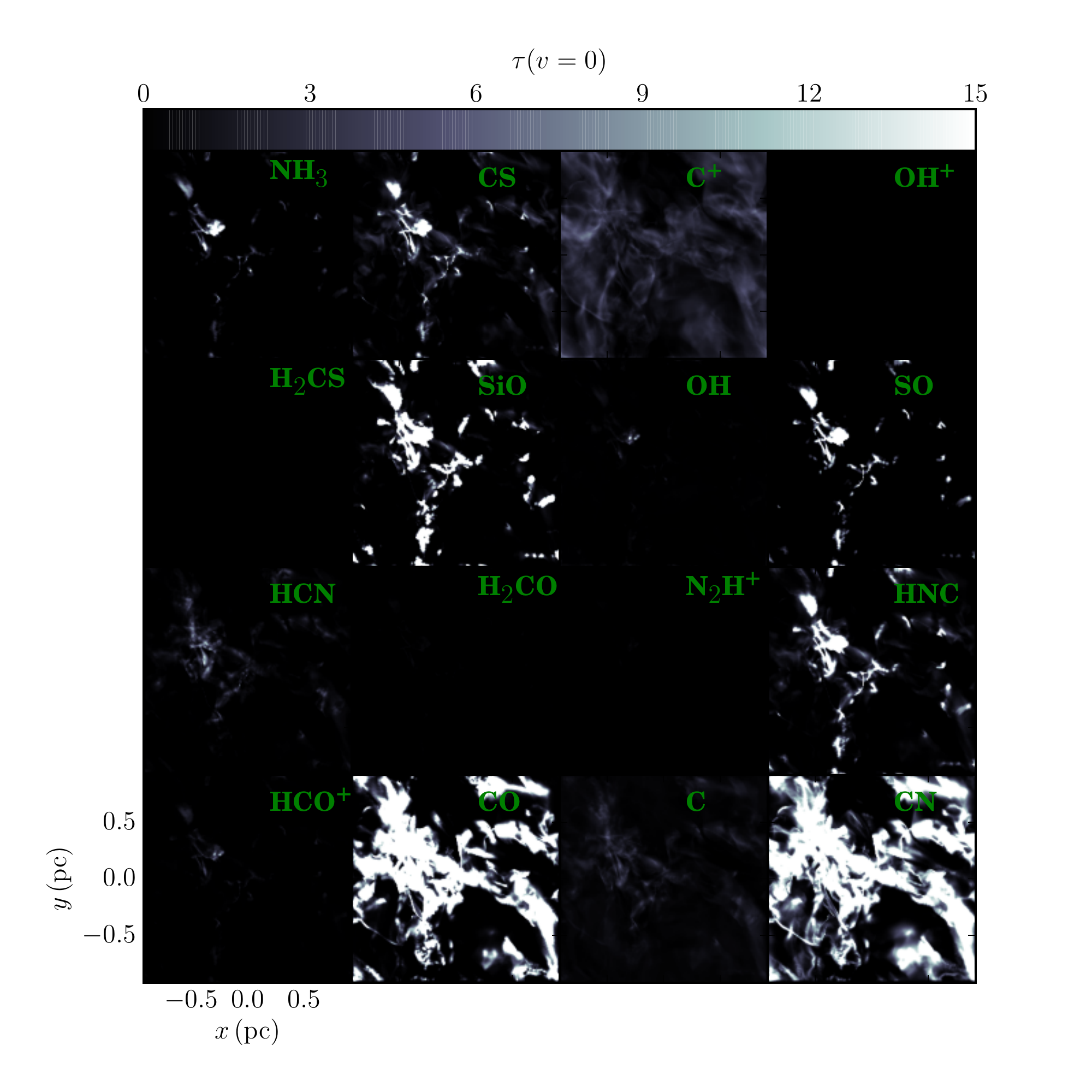}
\caption{\label{fig:tau} Line center optical depths calculated by {\sc radmc-3d}. The HCN and CO density profiles in this plot are treated with a crude dust treatment described in the Appendix. The HCN optical depth is divided by a factor of three to approximate the addition of the fine structure component to the transition.
}
\end{center}
\end{figure*}

In contrast, species with optically thin emission have very similar density and emission SCFs, as indicated by the darker cells in Figure \ref{fig:dOE}. For example, CO has a low critical density, $n_c \sim$ 2000 cm$^{-3}$, and a relatively high abundance. It becomes optically thick as the gas density approaches 10$^4$ cm$^{-3}$.  Figure ~\ref{fig:tau} demonstrates that CO is very optically thick throughout most of the simulation box. High optical depth effectively flattens the perceived distribution, i.e. as the gas becomes optically thick the emission no longer traces higher density regions. On the other hand, N$_2$H$^{+}$ is optically thin throughout the domain and has a small distance (d $<$ 0.1) between its emission and density SCF. N$_2$H$^+$ has a high critical density of 10$^5$ cm$^{-3}$ and has a much lower abundance than CO.

Differences between the true density and the emission arises from a combination of chemistry, which changes the abundance, and excitation, which impacts the line shape. For example, OH$^+$ has a low critical density of $\sim$ 4000 cm$^{-3}$ but its emission remains optically thin due to its very low abundance. However, HCN is optically thin because it is only excited at gas densities above 10$^6$ cm$^{-3}$ despite its modest  abundance: [HCN]/[H$_2$] $\sim$ 10$^{-7}$.

\begin{figure}[h!]
\begin{center}
\includegraphics[width=\columnwidth]{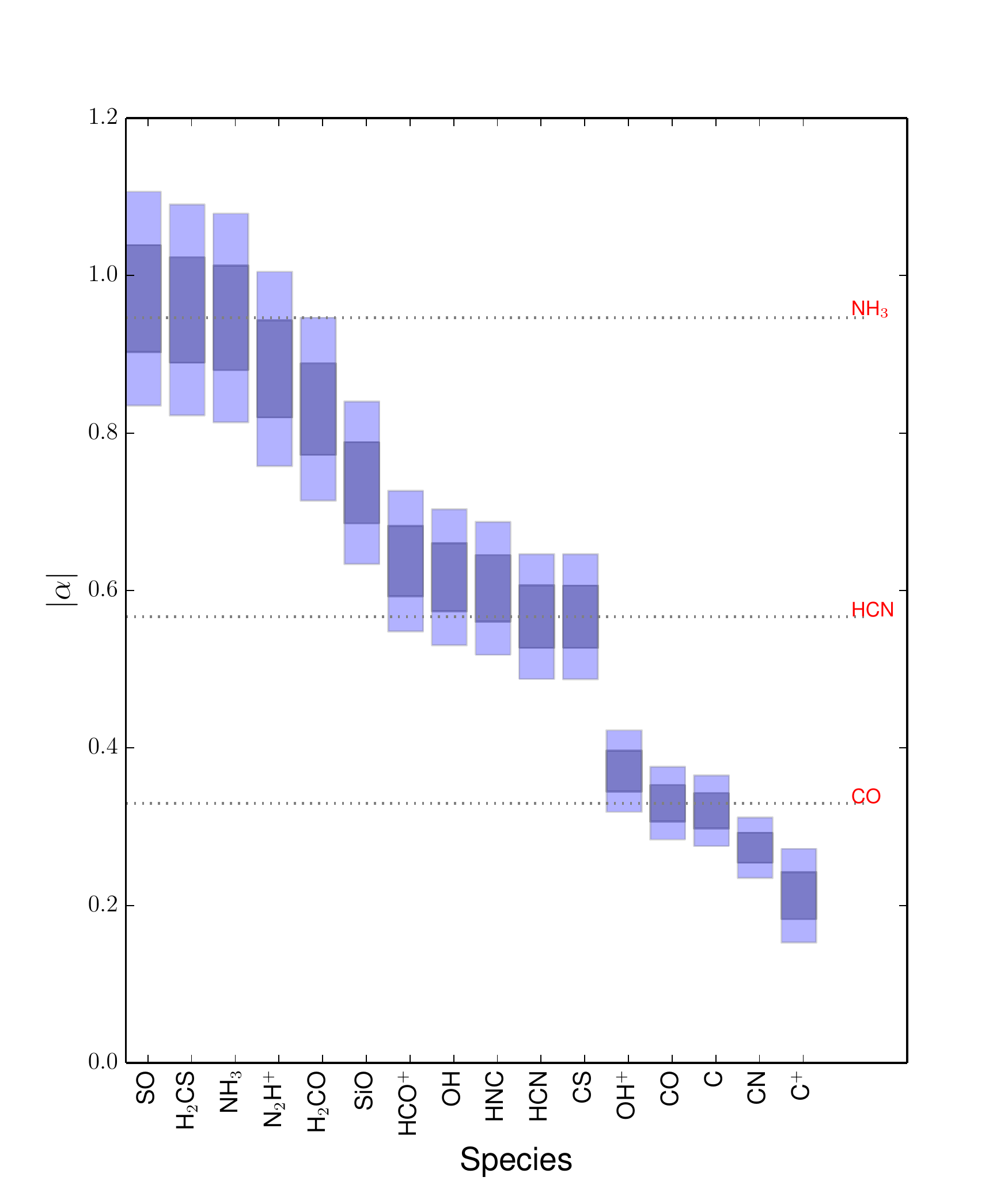}
\caption{ \label{fig:dalphC} Plot of the SCF slopes for all the species used in the study. The blue and gray boxes are the 1 and 2 $\sigma$ errors on the slope, respectively, from line of sight variations. The horizontal grey lines indicate the measured slopes for three common tracers.
}
\end{center}
\end{figure}

\begin{figure*}[h!]
\begin{center}
\includegraphics[width=0.8\textwidth]{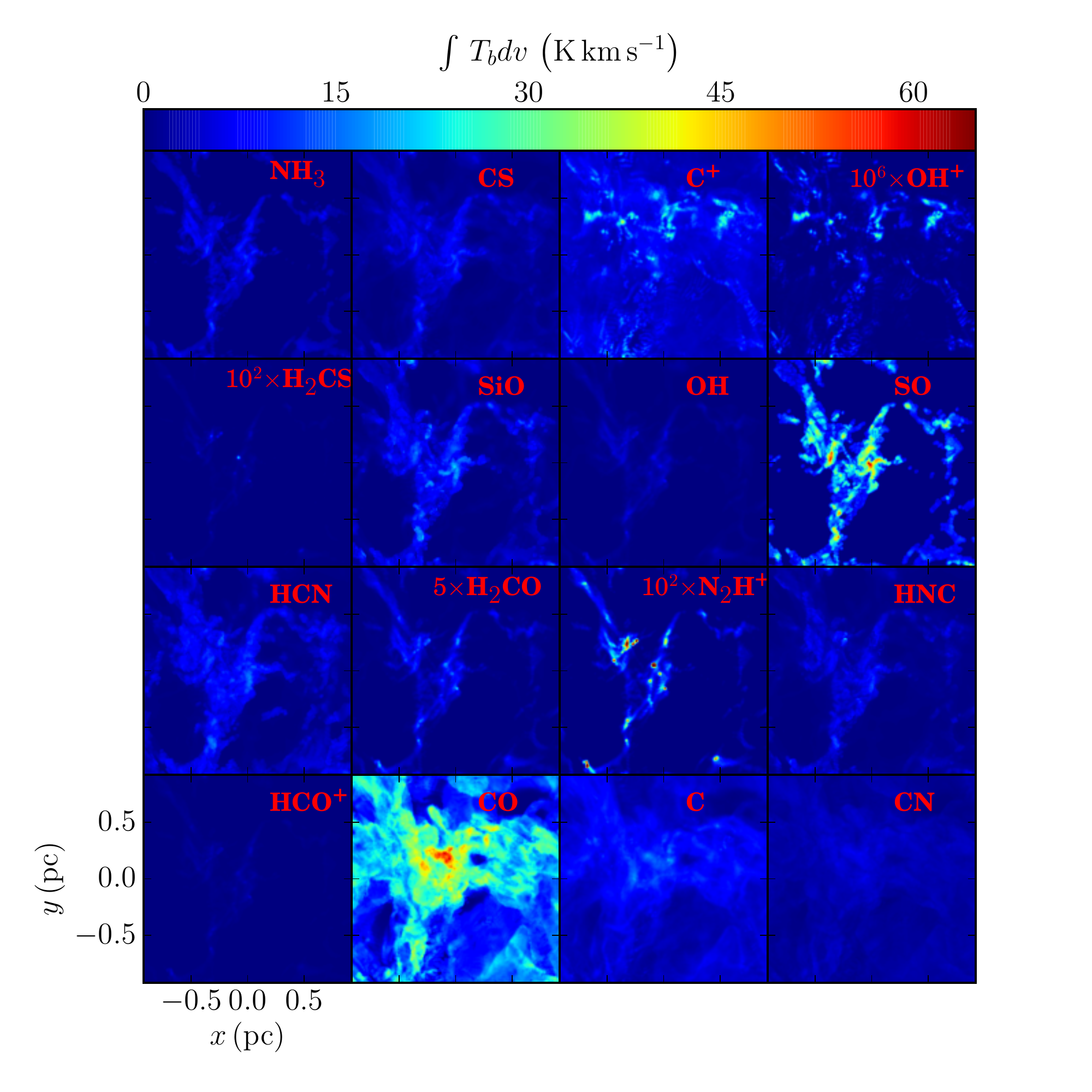}
\caption{\label{fig:xvsy} Integrated intensity maps for 16 different species, in units of $\rm{K \, km \, s^{-1}}$. The species OH$^{+}$, H$_2$CS, N$_2$H$^{+}$ and H$_2$CO have their emissions multiplied by the value shown  shown to see their structures.
}
\end{center}
\end{figure*}

\begin{figure*}[h!]
\begin{center}
\includegraphics[width=0.8\textwidth]{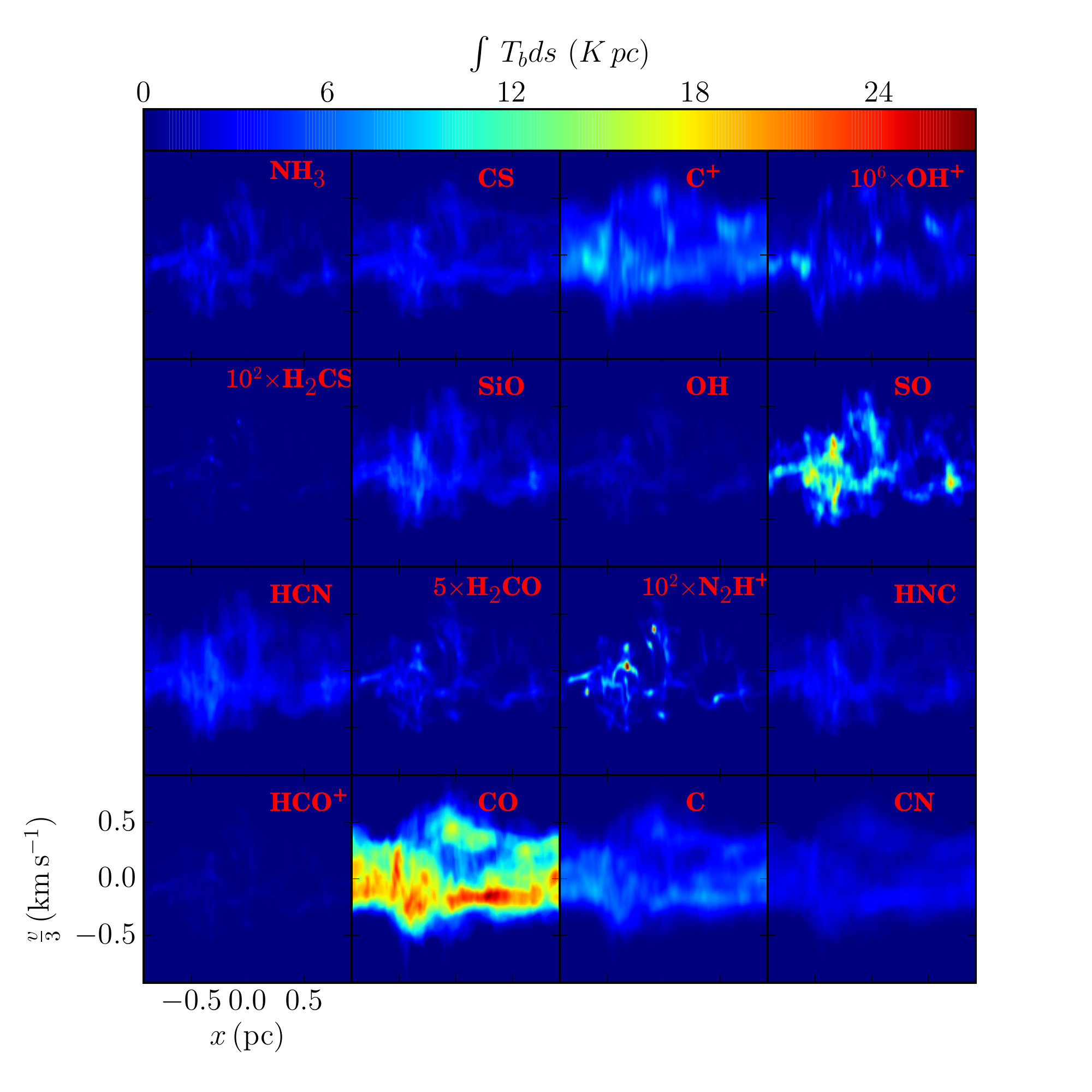}
\caption{\label{fig:xvsv} Integrated emission maps of velocity versus position, in units of $\rm{K \, pc}$. The velocity on the vertical axis is scaled by 3 $\rm{km s^{-1}}$. We multiply the emission of OH$^{+}$, H$_2$CS, N$_2$H$^{+}$ and H$_2$CO emission by the value shown to increase contrast.
}
\end{center}
\end{figure*}

\section{Discussion}
\subsection{Comparisons with Observations}\label{sec:compobs}
\begin{deluxetable}{ccc}
	\tablecolumns{3}
	\tablewidth{\columnwidth}
	\tablecaption{Comparison between Model and Observed Clouds \label{table:obs}}
	\tablehead{
		\colhead{Cloud} & \colhead{$\alpha$} & \colhead{$\sigma_{\rm fit}$}}
	\startdata
	 Model & -0.29  & 0.01\\
   	 Model (Oph Res) & -0.13 & 0.01\\
  	 Model (Per Res) & -0.19 & 0.006\\
  	 Ophiuchus & -0.22  & 0.01\\
 	 Perseus & -0.21 & 0.007 
	\enddata
	\tablecomments{Comparison of $^{12}$CO SCF power law fit values, $\alpha$. Per Res indicates the model cloud at the spatial resolution of the Perseus cloud, and likewise for Oph Res. $\sigma_{\rm fit}$ is the error of the power law fit in log space.}
\end{deluxetable}

\begin{figure}[h!]
\begin{center}
\includegraphics[width=\columnwidth]{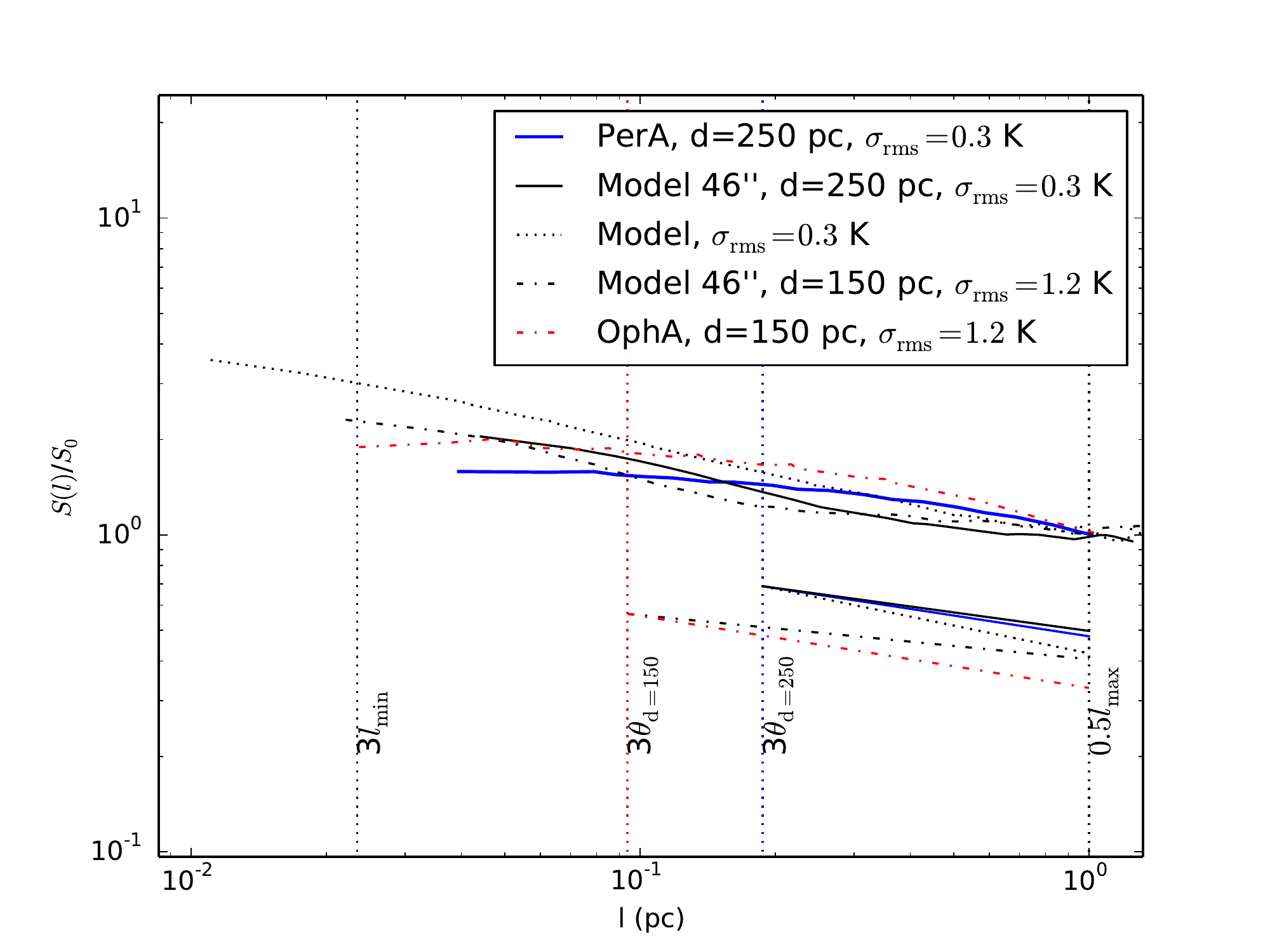}
\caption{\label{fig:comp}  SCF as a function of size for two clouds in the COMPLETE survey and the model cloud at different resolutions. The dashed-dot red line is the SCF for the Ophiuchus cloud, while the solid blue line is the Perseus cloud. We divide the Perseus cloud into two components at the cloud center (dashed blue lines). The different black line styles indicate different spatial resolutions, where the dotted line represents the simulation resolution, the solid line represents a 46'' beam at 250 pc, and the dot-dashed line represents 46'' at 125 pc to match with the line styles of the observed clouds. The vertical lines represent the minimum length scales used for the power law fitting, with $\theta$ representing the beam size. The power law fits are shown below, with the line styles matching their corresponding SCF.}
\end{center}
\end{figure}

We now compare our results to observations of two local molecular clouds (MCs), Perseus and Ophiuchus. We use the $^{12}$CO data from the COMPLETE survey \citep{2006AJ....131.2921R}, which were taken using the Five College Radio Astronomy Observatory (FCRAO). The $^{12}$CO observations have an angular resolution of 46$''$. Since the Perseus and Ophiuchus clouds are approximately 250 pc and 150 pc away \citep{2006AJ....131.2921R} corresponding to physical scales of 0.06 pc and 0.03 pc, respectively, we can compare at the to length scales represented in our simulation. We calculate the $^{12}$CO SCF for the full extent of both clouds. We also divide Perseus into two parts and compute the SCF of each half. Table ~\ref{table:obs} displays the resulting SCF slopes and power law fit errors. We plot the full SCFs in Figure ~\ref{fig:comp}. 

In order to properly compare the model to the observations, we perform several procedures on the observational data. First, we smooth the data with a Gaussian beam corresponding to the FCRAO 46$''$ beam resolution. Then, we regrid the data so that each pixel has the same spatial pixel scale of the two different clouds. Third, we apply a detection limit where we remove pixels with emission less than $0.01 \times <T_b>$. Finally, we add noise corresponding to Gaussian thermal noise with a standard deviation of $\sigma_T \sim $ 0.3 K to model the noise in the Perseus COMPLETE data. We did not perform biased sampling, which may affect the longest length scales of the SCF due to the irregular observational stencil.

Table \ref{table:obs} gives the values of the slopes and the associated error of the power law fit. The model cloud SCF initially has a slope of -0.29. However, after matching to the observational resolution the SCF slope changes by quite a bit to -0.13 and -0.19 for the Ophiuchus and Perseus resolutions, respectively. Furthermore, the fit error indicated in the table shows that the power law fits are all well constrained. Figure ~\ref{fig:comp} shows that the model cloud and the Perseus cloud have similar SCFs when taking into account observational biases, such as beam size and pixel resolution. This is indicated by the similar power law fits shown in the bottom of the figure. The Ophiuchus SCF is steeper than the mode SCF at the Ophiuchus pixel and beam resolution, though it is similar to both the Perseus SCF and the model SCF at Perseus resolution. The flattening of the SCF at small scales is due to the beam resolution. While the emission is still correlated, the turbulence has been resolved out within the beam area. There is still some difference between the simulation and observations which may be due to other biases. However the model and observations match within the 10\% error mentioned in \S\ref{sec:va}.

In Figure \ref{fig:scfs}, the SCFs for all of the tracers appear to flatten out at some large scale, with the exact scale seemingly changing depending on the tracer. \citet{2003ApJ...588..881P} also found that their SCFs from simulated clouds traced by $^{13}$CO flatten, but the SCFs for local clouds, also traced by $^{13}$CO, steepen. We were not able to replicate this discrepancy with our simulations, though we conclude that the steepening effect is most likely due to some observational bias at large length scales.

\subsection{Understanding Chemistry from the SCF}\label{sec:understand_chem}

A fundamental question of this work is: what does the SCF reveal about the chemistry of various tracers? Using Figures \ref{fig:sn} and \ref{fig:dalphC} we group the SCF slopes into three emission categories: diffuse, intermediate, and compact. C is an example diffuse tracer which is excited down to lower densities of $\sim$ 800 cm$^{-3}$ and is abundant throughout the entire volume. HCN is excited at fairly high densities and it becomes quite abundant once it is shielded from the UV field, so it traces up to intermediate scales. A common dense core tracer is N$_2$H$^+$ which also gets excited at higher densities, where it is also shielded from photodissociation. Commonly studied tracers such as CO, C$^{+}$, HCN, NH$_3$ and N$_2$H$^{+}$ confirm that the SCF slope reflects the expected optical depth, filling fraction, and critical density of the emission. While it is impossible to compute the true density distribution from the SCF alone, the slope does indicate the scale the species traces. For instance, shallower slopes, such as that of CO, show that the gas remains correlated on larger scales. 

CO is a species of particular astrophysical importance. Over the past several decades, CO has become the most prominently utilized cloud mass tracer and has received significant theoretical attention. Our results show that the SCF of CO traces diffuse regions, as indicated by its shallow slope. Our results also show that CO traces the gas up to high densities, although proper treatment of dust grain depletion could change this. Since CO has a low critical density (see Table \ref{table:trans}), the lowest level transition is easily excited throughout most of the cloud. This allows the cloud to produce emission in both the lower density environments and the dense cores (where the emission saturates due to optical depth). For this reason, CO also has a high volume and surface filling fraction. Figure \ref{fig:tau} confirms that CO has the highest average optical depth, which is due to its low critical density and higher abundance. 


However, the SCF analysis identifies several tracers that exhibit similar emission. We find that C and CN are homologous to CO. The correlations between these species makes chemical sense since all three depend upon the abundance of neutral Carbon. The UMIST chemical network shows that both CO and CN form rapidly in the original diffuse environment. Their behavior at higher densities though is quite different. CN is photodissociated much more rapidly towards lower densities than CO because it cannot self shield. CN is depleted through reactions forming more complex molecules faster than CO. This results in a factor of several orders of magnitude difference between the abundances of the two molecules. Therefore, it is reasonable to expect CO and CN to be very similar in low density environments. In regions of high mass star formation, CN could be used as a proxy for CO, which is optically thick, or of the surrounding diffuse gas.

Another common molecular gas tracer is HCN, which is frequently used to trace gas with densities above 10$^6$ cm$^{-3}$. The SCF slope implies it also traces intermediate density regions in filaments, i.e. $n \sim$ 10$^3$ to 10$^4$ cm$^{-3}$, although, the emission level may be too low to detect. Recent work by \citet{2014arXiv1406.0540F} confirms that HCN and HNC are good tracers of dense environments over a wide range of extinctions. The optical depths we calculate for HCN are off by a factor of three since we do not model the HCN fine structure for the ground state transition.

Finally, there are a variety of species that trace dense cores, including NH$_3$ and N$_2$H$^{+}$.  Figure ~\ref{fig:xvsv} confirms the trend that species tracing more diffuse regions have shallower slopes than species that typically trace high density regions. Here, the regions the species trace are clear in the velocity information. Diffuse gas tracers have significant emission in a broad range of velocities, as illustrated by the horizontal bands in species CO, C and CN in Figure \ref{fig:xvsv}. Species which trace higher densities show no diffuse component. Instead, the emission is located in clumps or filaments, which exhibit a smaller range of velocities. 

While the density is important, the gas must also have a high enough temperature to excite the  transition. We define a characteristic line temperature T, where $h\nu_i = k T_{i}$, where $\nu_i$ is the line frequency. All the species have line temperatures below 10 K except C, C$^+$ and OH$^+$, which have line temperatures of 23 K, 91 K and 44 K, respectively. The average gas temperature in diffuse regions is around 100 K, so even these species are easily excited from their ground states. However, OH$^+$ is mostly observed in absorption. This is due to its low abundance and higher excitation making measurements from absorption in the dust continuum easier than trying to detect the very small emission signal (as noted by the multiplicative factor of $\sim$ 10$^6$ in Figures \ref{fig:xvsy} and \ref{fig:xvsv}).

Complex molecules, such as NH$_3$ require higher densities to form, and photodissociate rapidly at lower densities where the UV field is higher. Simple light diatomic molecules such as CN form in diffuse regions. An exception is CS, which appears only in intermediate density regions. It photodissociates faster than CO but slower then CN. However, due to the much lower initial abundance of sulfur, it forms only where the sulfur is concentrated.  Most of the tracers in this study that have a shallow slope also tend to have a very high optical depth. The only exception is OH$^{+}$ which has a very low abundance. 

\subsection{Discussion of Observational Implications}\label{sec:implic}
The results from Figures ~\ref{fig:sn} and ~\ref{fig:dE}, as well as our definition of complementary species, suggest sets of homologous species. These are groups of species whose spectral structure is very similar in PPV space, indicating that, especially for the optically thin tracers, they should trace similar density regimes.

Recent theoretical studies, such as \citet{2014arXiv1403.3530G} and \citet{2014MNRAS.tmpL..37O}, found that C is a good alternative tracer to CO. C has several advantages, including a ground state transition at 609 $\rm{\mu m}$ and a lower optical depth than CO. These recent studies challenge the idea that C traces only the surface of the PDR. Our study confirms this picture by showing that C and CO have very similar SCFs, with CO being complementary to C. Our study {\it also} predicts that CN is an alternative tracer to CO and C. CN (1-0) has a similar transition frequency with an optical depth around an order of magnitude less than CO, although its slope is slightly flatter indicating that it traces even lower density regions, and it has a similarly high filling fraction. The slightly flatter slope is expected; CN is destroyed faster than CO in higher density environments due to its role as a reactant in reactions forming more complex molecules. Since CN and CO form very quickly and depend on the relatively high-abundance of C, they both have very high filling fractions. Their surface filling fractions are both over 90\%, and their volume filling fractions are greater than 0.3. In fact, Table \ref{table:trans} shows that CN is both more surface filling and volume filling than CO. The cosmic-ray induced photodissociation rates (i.e. CX + CRPHOT $\rightarrow$ C + X) given by the UMIST2012 network are R$_{\rm{CN}} = 1.4 \times 10^{-13}$ { s$^{-1}$} and R$_{\rm{CO}} = 7.5 \times 10^{-16}$ { s$^{-1}$}. Cosmic rays penetrate further into the cloud than the external UV radiation, indicating the CN will be destroyed by cosmic rays inside the cloud faster than CO. Indeed, CO has an abundance several orders of magnitude greater. Singly ionized carbon traces more diffuse regions but is not as abundant in higher density regions where it combines with O to form CO. The C$^{+}$ (1-0) transition is in the infrared, which can be observed using space-based instruments, such as the Herschel Space Observatory \citep[e.g.][]{2014A&A...561A.122L} and the GREAT Spectrometer on Stratospheric Observatory for Infrared Astronomy (SOFIA).

A commonly used high density tracer is N$_2$H$^{+}$, which has a ground state transition at 93 GHz. Our study predicts several homologous tracers to N$_2$H$^+$ including: H$_2$CO, H$_2$CS, NH$_3$, and SO. All of these species exist in similar environments and have surface filling fractions between 0.01 and 0.25. However, some of these tracers only form the high density cores which such as N$_2$H$^+$, with a volume filling fraction of 0.21. H$_2$CO has a higher volume filling fraction than N$_2$H$^{+}$ (f$_v$ = 0.37) which indicates that it traces a larger fraction of the core gas. We find that  H$_2$CO and SO have higher brightness temperatures indicating that they should be easier to detect than species with fainter emission such as H$_2$CS and N$_2$H$^+$.  NH$_3$ and SO show significant emission in filaments, but N$_2$H$^{+}$ is brightest in dense ``cores". NH$_3$ has a relatively low critical density (n$_c = 1991$ cm$^{-3}$), so it is excited at lower densities than high densities tracers like N$_2$H$^+$. However, it is not as bright as other low critical density tracers like CO ($n_c \sim 2000$ cm$^{-3}$) because of its low abundance outside of dense regions.  Some of these correlations could change with the inclusion of dust grain chemistry. This would lower the abundances of the higher density tracers, such as NH$_3$, H$_2$CO, H$_2$CS and SO, and reduce their emission. However, these molecules only begin to deplete for H$_2$ number densities $\geq$ 10$^7$, greater than the maximum density in this simulation. Detailed treatment of gas-grain chemistry is beyond the scope of this work.

\section{Conclusions}
We use numerical simulations of a Milky Way-like molecular cloud to study how the SCF varies between different species. We post process the hydrodynamical simulation with the astrochemistry code {\sc 3d-pdr}, which uses a full chemical network to obtain abundances of over 200 different species. We produce synthetic line observations for a subset of these. We calculate the SCF for each of the species for both the density and emission distributions and define a ``distance" metric to compare them. On the basis of this, we draw the following conclusions:

\begin{enumerate}
    \item The SCF is sensitive to the chemistry of the tracer. Species tracing diffuse gas tend to have shallower SCF slopes, whereas species that trace dense regions ("cores") have steeper slopes. 
    \item We confirm that the emission structure, as characterized by the SCF, poorly traces the density structure for species with optically thick emission. The decoupling is due to line saturation in the highest density regions.
    \item Spatial resolution has a distinct effect on the SCF slope, but even with relatively low beam resolution the slopes remain in a similar region of parameter space. However, for poor resolution, some species have artificially similar SCFs.
    \item Velocity resolution has no effect on the SCF slope, since the SCF only measures the rms velocity between spatial regions. This will only hold while the spectral features are resolved, namely that the velocity channels are smaller than the linewidth. Noise variation also has little effect on the slope, since variations cancel as long as the noise is Gaussian.
    \item  We find that C, C$^{+}$, CN, CO, and OH$^{+}$ are homologous diffuse gas tracers. OH, CS, HCN, HNC and SiO are homologous intermediate gas tracers. Finally, H$_2$CO, H$_2$CS, N$_2$H$^{+}$ and NH$_3$ are homologous dense core tracers when gas phase chemistry is dominant. The statistical similarity of the SCFs suggest that they trace similar cloud structure and likely provide complementary information.
\end{enumerate}
This study takes the first steps in exploring the 3D astrochemical correlations in molecular clouds. In this study, we show that the SCF slope can be used as an indicator of the density of environments where specific species form. This provides insight into the gas chemistry of particular species. Future work can still expand on this in several ways. We do not investigate the SCF evolution as a function of time. Also, we do not include either dust grain chemistry or shock chemistry. Future studies can investigate higher level transitions, such as CS (2-1) and CO (3-2), and isotopologues such as $^{13}$CO.

\appendix
\section{Chemistry Considerations}
In this appendix, we discuss briefly how effects such as dust-grain chemistry and shock chemistry could affect our results. 
\subsection{CO and HCN}
Both CO and HCN, as well as other high-density tracers, may be affected by depletion at high-densities. CO depletion is extreme at high-densities with the mean abundances declining by several orders of magnitude \citep[see][]{2002A&A...389L...6B, 2007ARA&A..45..339B, 2014MNRAS.438L..56H}. HCN also freezes out, but at higher densities. In order to assess the impact of depletion on our CO and HCN results, we adopt crude treatments for dust freeze-out and recalculate the optical depth and SCF. We set the abundance of CO to 0 where the H$_2$ number density exceeds 10$^4$ cm$^{-3}$. For HCN we set a limit of [HCN]/[H$_2$] $=$ 10$^{-8}$ in cells where $n_{\rm H_2} \geq 10^4$ cm$^{-3}$. We compare the SCF slopes of the tracers with and without depletion and find that the change in slope is less than 10\%.  Setting a maximum HCN abundance of 10$^{-8}$ relative to H$_2$ reduced the HCN optical depths to $\tau < 10$ across the entire cube, with the typical optical depth being $\sim$ 1-4.  Furthermore, since in these high densities the HCN gas will remain optically thick, it should not drastically impact the SCF correlations. Overall, we conclude that the exclusion of dust depletion does not significantly affect our results due to the small percentage of the simulation volume at high-densities (1\%). 
\subsection{N$_2$H$^+$}
When CO starts to freeze-out onto dust grains, several molecules, such as N$_2$H$^+$ will see an increase in their abundances. This occurs because H$_3^+$ reacts with CO to form HCO$^+$ which is the main destruction mechanism for H$_3^+$. As the amount of CO decreases there is a surplus of H$_3^+$. In high density environments, H$_3^+$ can be destroyed to form N$_2$H$^+$ \citep[][see]{2001ApJ...552..639A, 2002ApJ...570L.101B,2004MNRAS.352..600R} by the following mechanism:
\begin{equation}
\textrm{H}_3^+ + \textrm{N}_2 \rightarrow \textrm{N}_2\textrm{H}^+ + \textrm{H}_2
\end{equation}
In order to test whether this effect has an impact, we adopt a simple approximation in which the amount of N$_2$H$^+$ increases at high densities. Since CO starts to deplete at H$_2$ densities above n$(H_2) \ge 10^4$ cm$^{-3}$, we multiply the N$_2$H$^+$ abundance by a factor of 4 where the gas fits this criteria. Since the emission is mostly optically thin, this merely scales the emission by a constant factor. Similar to the case for CO and HCN depletion, we find that this mechanism has no statistical impact on our results. It is noteworthy that the enhancement of N$_2$H$^+$ is not universal, with studies such as \citet{2002ApJ...569..815T, 2004A&A...416..191T} showing either no N$_2$H$^+$ increase or showing N$_2$H$^+$ depletion.
\subsection{Shock Chemistry: SiO}
Since our hydrodynamic simulation doesn't resolve shock fronts to the necessary resolution to calculate post-shock densities and temperatures, well known shock tracers such as SiO will not be properly modelled. In shocks, changes in the density and temperature can lead to a large enhancement of the abundance of SiO. Furthermore, at higher temperatures, the abundance of SiO can be increased by sputtering from dust grains. This mechanism is the ejection of Si and SiO from a grain surface following a high enough energy impact of gaseous species. At lower temperatures, the sputtering rate is expected to be small \citep[][]{1979ApJ...231...77D}. Even though sputtering from shock chemistry, and other shock effects, will affect the abundance, the inclusion of shock chemistry is beyond the scope of this work. Finally, the increased brightness of the SiO emission is an artifact of using a non-depleted Si abundance, which is over an order of magnitude larger that observed depleted abundances \citep[][]{1975ApJ...197...85M}.

\acknowledgments
The authors acknowledge support from NASA through Hubble Fellowship grant
\# 51311.01 awarded by the Space Telescope Science Institute, which is operated by the Association of Universities for Research in Astronomy, INC., for NASA, under contract NAS 5-26555 (SSRO). ER is supported by a Discovery Grant from NSERC of Canada. TGB acknowledges the NORDITA programme on Photo-Evaporation in Astrophysical Systems (2013 June). The ORION and 3D-PDR calculations were per formed on the Trestles XSEDE cluster and DiRAC-II COSMOS supercomputer (ST/J005673/1, ST/H008586/1, ST/K00333X/1), respectively. The authors thank R. Snell, N. Evans and K. \"{O}berg and the anonymous referee for their valuable comments and suggestions that greatly improved this work.

\bibliography{converted_to_latex.bib}

\end{document}